\documentclass[aps,prb,twocolumn,groupedaddress,superscriptaddress,showpacs,floatfix,longbibliography]{revtex4-1}
\usepackage{graphicx}
\usepackage{physics}

\usepackage{xcolor}
\usepackage{amsfonts}
\usepackage{amsmath}
\usepackage{amssymb}
\usepackage{rotating}
\usepackage{bm}
\usepackage{lipsum}
\usepackage{bbm}
\usepackage{kantlipsum}
\usepackage{braket}
\usepackage{dsfont}

\def\be{\begin{eqnarray}}
\def\ee{\end{eqnarray}}

\usepackage[unicode=true, bookmarks=false,breaklinks=false,pdfborder={0 0 1},backref=false,colorlinks=false]{hyperref}
\hypersetup{hypertex}   % gera link para referencias ao longo do texto
\usepackage{breakurl}
\hypersetup{backref,colorlinks=true,citecolor=blue,linkcolor=blue,filecolor=blue,urlcolor=blue}
\usepackage[english]{babel}

\begin{document}

\title{%Orbital orientation as a design principles for the Rashba effect
Spin-deformation coupling in two-dimensional polar materials
%Rashba spin-orbit coupling and non-Abelian gauge invariance: The absence of spin-deformation coupling in curved surfaces
}
%Graphene flat bands driven by he strong light-matter coupling

\author{J. A. Sánchez-Monroy}
\affiliation{Departamento de Física, Universidad Nacional de Colombia, Bogotá, 111321, Colombia}
\email[]{jasanchezm@unal.edu.co}

\author{Carlos Mera Acosta}
\email[]{carlos.mera@unisabana.edu.co}
\affiliation{Faculty of Engineering, University of La Sabana, Chia 250001, Colombia}
\affiliation{Center for Natural and Human Sciences, Federal University of ABC, Santo Andre, SP, Brazil}

%\date{\today}

\begin{abstract}
%In spintronics, controlling the spin via spin-orbit coupling (SOC), magnetism, or topological effects is key. This paper discusses spin-deformation coupling (SDC) as another control mechanism, focusing on polar-deformed films or 2D compounds in which the Rashba SOC is introduced as an $SU(2)$ gauge field. Using the thin-layer approach and a suitable gauge transformation, we decoupled the surface and the normal electronic degrees of freedom. We find that: \textit{i}) gauge invariance implies that the spin is uncoupled from the surface's extrinsic geometry, challenging the common consensus, and \textit{ii}) apart from the spin-orbit coupling arising from the $SU(2)$ gauge field in curvilinear coordinates, we uncover a scalar potential dependent on the Rashba SOC strength. The outcomes of our work open novel pathways for exploring the manipulation of spin degrees of freedom through the use of the SDC. 

The control of the spin degree of freedom is at the heart of spintronics, which can potentially be achieved by spin-orbit coupling or band topological effects. In this paper, we explore another potential controlled mechanism under debate: the spin-deformation coupling (SDC) - the coupling between intrinsic or extrinsic geometrical deformations and the spin degree of freedom. We focus on polar-deformed thin films or two-dimensional compounds, where the Rashba spin-orbit coupling (SOC) is considered as an $SU(2)$ non-Abelian gauge field. We demonstrate that the dynamics between surface and normal electronic degrees of freedom can be properly decoupled using the thin-layer approach by performing a suitable gauge transformation, as introduced in the context of many-body correlated systems. Our work leads to three significant results: (i) gauge invariance implies that the spin is uncoupled from the surface's extrinsic geometry, challenging the common consensus; (ii) the Rashba SOC on a curved surface can be included as an $SU(2)$ non-Abelian gauge field in curvilinear coordinates; and (iii) we identify a previously unnoticed scalar geometrical potential dependent on the Rashba SOC strength. This scalar potential, independent of spin, represents the residual effect remaining after decoupling the normal component of the non-Abelian gauge field. The outcomes of our work open novel pathways for exploring the manipulation of spin degrees of freedom through the use of the SDC.

\end{abstract}

\maketitle

%In order to describe the spin-deformation coupling, we illustrate the perturvative expansion for \textit{i}) the of phenomenological linear Rashba effect and \textit{i})a non-Abelian gauge field capturing all spin effects (including the linear Rashba SOC).   

\section{Introduction}

Device applications not only require an understanding of the physical mechanisms behind material properties, but also the ability to control or modify them. In the particular case of spintronics, which focuses on manipulating spin-related properties in materials, numerous mechanisms have been proposed to achieve spin polarization (SP) control~\cite{PhysRevLett.122.036401,science.1184028,RevModPhys.76.323,PhysRevApplied.4.047001}, including band topological effects~\cite{PhysRevLett.112.076802,Costa2021-hb}, magnetoelectric effects~\cite{Kimura2003-gz,annurev-matsci-070813-113315,Bea2008-ol}, and spin-orbit coupling (SOC)\cite{PhysRevB.94.041302,Mera_Acosta2019-pn,PhysRevB.102.144106,Long2018-br,He2022-wd,Smejkal2018-pw}. One of the most studied SOC-driven phenomena is the SP induced by the breaking of the inversion symmetry, e.g., the Rashba effect in polar compounds~\cite{MERAACOSTA2020145,Manchon2015-lp,Nascimento2022-bi}.
%in compounds that are polar, non-polar chiral, and non-polar non-chiral, which are usually referred as bulk Rashba, Weyl, and Dresselhaus effects, respectively. 
In general, the SP directions are determined by the local symmetry of the chemical environment for the constituent atoms and the crystallographic point group symmetries~\cite{PhysRevB.104.104408}.
Consequently, geometric deformations hold the potential to modulate SP and hence, it offers a route to manipulate spin related properties. 
The underlying assumption of such a SP control mechanism in nanodevices is the existence of the coupling between the spin degree of freedom and local geometrical deformations~\cite{Chang2013a,Wang2017a,Taichi2011,Siu2018,Zhang2007a,Liu2011a,Ortix2015ab,Gentile2015a,Ying2016a,Pandey2018a,Salamone2022a}, i.e., \textit{spin-deformation coupling} (SDC). The theoretical modelling of the SDC requires having a Hamiltonian that describes the electronic structure of electrons confined to a deformed surface, denoted as $\mathcal{M}$, where surface modes are decoupled from normal degrees of freedom. 
Two dimensional (2D) materials and surfaces are thus a platform to induce deformations and potentially realize the SDC.

The SDC can be described by the coupling of spin with two distinct manifestations of the deformation, namely: \textit{i}) the \textit{intrinsic SDC} - geometrical attributes that exclusively involve the intrinsic geometry of the surface, devoid of any influence from the embedding space, e.g., the induced metric tensor ($g_{\mu \nu}$) and the Gaussian curvature ($\text{K}$), and \textit{ii}) the \textit{extrinsic SDC} - the mathematical constructs depending on the extrinsic geometry, involving the curvature of a surface within its 3D spatial embedding, such as the mean curvature ($\text{H}$). Diverse methodologies have been proposed to model the SDC on curved surfaces \cite{Chang2013a,Wang2017a,Taichi2011,Siu2018} and in quantum wires \cite{Zhang2007a,Liu2011a,Ortix2015ab,Gentile2015a,Ying2016a,Pandey2018a,Salamone2022a}.
For instance, concerning curved two-dimensional Rashba compounds, the prevailing consensus establishes a SDC involving the coupling of the spin with the extrinsic curvature, which is typically associated with a term proportional to the product involving $\text{H}$ and a spin component \cite{Chang2013a,Wang2017a,Taichi2011,Siu2018}, i.e., $\xi\text{H}\tilde{\sigma}^i$, where $\tilde{\sigma}^i$ denotes the spatially dependent Pauli matrices, and $\xi$ represents the strength of the coupling. For the intrinsic SDC in Rashba compounds, the electronic Hamiltonian incorporates terms coupling the metric on the surface with spin components \cite{Chang2013a,Wang2017a,Taichi2011,Siu2018}.

%In these treatments, it is usually addressed using the thin-layer method, in which a particle is confined to the surface (or curve) by a strong potential. 
%However, as we demonstrate in this work, the coupling between the spin and the mean curvature $\xi\text{H}\tilde{\sigma}^i$ is only one aspect of the SDC. For instance, the SOC could also couple to other attributes of the curvature such as the metric on the surface or the Gaussian curvature.  Specifically, as detailed below, we describe the SDC in terms of three contributions: \textit{i}) the Rashba SOC written in the curved surface that involves the product of the surface metric and the space-dependent Pauli matrices, \textit{ii}) the usual reported SDC, i.e., the coupling between the spin and the mean curvature, and \textit{iii}) the SOC-induced modulation of the deformation, i.e., the product between SOC strength and an attribute of the intrinsic or extrinsic deformation. 
%When SOC is included, the SDC emerges as a term proportional to the products between $\text{H}$ and a spin part, e.g., $\mathcal{H}_{SDC}=\xi\text{H}\tilde{\sigma}^i$, where the $\tilde{\sigma}^i$ are the space-dependent Pauli matrices and $\xi$ is the strength of the coupling.

The explorations of the SDC share a common foundation: the thin-layer method. This method has been widely utilized to describe non-relativistic spinless charged particles~\cite{Ferrari2008,Brandt_2017}, relativistic spinless particles \cite{Brandt_2017}, and $\frac{1}{2}$-spin particles \cite{monroy2017teoria} moving on curved surfaces in the presence of an external electromagnetic field; in all these situations, the coupling between electromagnetic fields and the extrinsic curvature is forbidden. 
Interestingly, when the procedure is not executed accurately, such as through improper decoupling of degrees of freedom~\cite{Jensen2009a}, a fictitious field-deformation coupling resembling the term $\xi\text{H}\tilde{\sigma}^i$ can emerge. This coupling has been identified as an unphysical gauge-dependent contribution, as demonstrated in~\cite{Ferrari2008,Ortix2011}. The close analogy between the SOC and electromagnetic fields suggests exploring the SDC by employing the well-established interpretation of the Rashba SOC as an effective non-Abelian field \cite{Frohlich1993,Jin_2006,Tokatly2008}, which is the purpose of this paper.

In this paper, we describe the SDC in the context of two descriptions of the spin polarization in 2D non-centrosymmetric materials, namely, \textit{i}) the phenomenological linear Rashba effect, i.e., $\mathcal{H}_{R}=\alpha_{R}\left(\sigma_{x}k_{y}-\sigma_{y}k_{x}\right)$, where $\alpha_{R}$ is the Rashba SOC strength, typically referred to as Rashba parameter; and \textit{ii}) a non-Abelian gauge field capturing all spin effects (including the linear Rashba SOC). In order to describe the SDC, we first discuss the metric tensor of deformed surfaces in the context of the thin-layer procedure. Subsequently, we deduce the SDC based on the two aforementioned types of SOC descriptions. 
%\textcolor{red}{
Writing the SOC as a $SU(2)$ non-Abelian  gauge field $\boldsymbol{\mathcal{W}}$ %that also captured the Zeeman coupling 
and using the thin-layer method \cite{Costa1981,Maraner95,Jaffe2003}, we demonstrate that the extrinsic SDC proportional to $\text{H}\tilde{\sigma}^i$  is physically forbidden due to gauge invariance. 
Indeed, only intrinsic SDC can be used as a SP control mechanism. We identify a previously unnoticed scalar geometrical spin-dependent potential proportional to the Rashba SOC strength. This represents a shift induced by the non-Abelian gauge field's normal component. 
%Further details of the general case encompassing curved surfaces and wires are provided in the supplementary material.}
%we identified terms that could contribute to achieving SDC. 
%Finally, we discuss the relation between the SDC and the coupling between electromagnetic fields with sruface deformations, and the necessary conditions to obtain a possible SDC.
%The explorations of SP control through the SDC, $\mathcal{H}_{SDC}=\xi\text{H}\tilde{\sigma}^i$, share a common foundation: the thin-layer method.

In order to illustrate the proposed description of the SDC, in the next section we describe the formalism to describe a curved surface in the context of the thin-layer method applied to the kinetic energy and confinement potential. In sections III and IV, the SDC is studied describing the Rashba SOC as a linear phenomenological model and a non-Abelian gauge field, respectively. In sections V and VI, we describe the controllability of the SDC and give a specific example of the proposed formalism. Finally, conclusions are given in section VII.  

%Curiously, a fictitious field-deformation coupling, resembling the term $\xi\text{H}\tilde{\sigma}^i$, can emerges \cite{Jensen2009a} when the procedure may not have been executed accurately, e.g., a proper choice of gauge~\cite{Ferrari2008,Ortix2011}.
%This contrast could be interpreted as consequence of some missed details in the treatment of the decoupling procedure. 
%%%%%%%%%%%%%%%%%%%%%%%%%%%%%%%%%%%%%%%%%%%%%%%%%%%%
\section{Spinless electrons confined in a deformed two-dimensional surface}
%%%%%%%%%%%%%%%%%%%%%%%%%%%%%%%%%%%%%%%%%%%%%%%%%%%%
%\textcolor{red}{RECIPE}
The general procedure for describing the SDC consists of identifying the coupling between spin and both the intrinsic and extrinsic curvatures in the effective Hamiltonian describing a deformed low-dimensional material\cite{Costa1981,Jaffe2003}. 
Consequently, the problem is translated to include the geometrical deformation in the Hamiltonian, constrained by confinement within the 2D space. 
To achieve this goal, we define the deformation of interest in terms of a metric tensor, and then apply the thin-layer method to the electronic Hamiltonian $\mathcal{H}_{e}$. For illustrative purposes, we first focus on the spinless electronic Hamiltonian.

%####################################
\subsection{Geometrical deformation parametrization} A local geometric deformation is introduced by considering the electronic positions on a curved surface $\mathcal{M}$, which is parameterized by $\boldsymbol{r}_{||}=\boldsymbol{r}(q_{1},q_{2})$. 
In a sufficiently small neighborhood of $\mathcal{M}$, the position vector can be described using the normal coordinate $q_{3}$, as $\boldsymbol{R}(q_{1},q_{2},q_{3})=\boldsymbol{r}_{||}+q_3\hat{\boldsymbol{n}}$, where $\hat{\boldsymbol{n}}(q_1,q_2)$ is a vector field normal to the surface $\mathcal{M}$. The transformed positions from the Cartesian coordinate system to the adapted curvilinear coordinates, $(q_{1},q_{2}, q_{3})$, define the metric tensor 
\begin{equation}
G_{ij}=\frac{\partial \boldsymbol{R}}{\partial q^i}\cdot\frac{\partial \boldsymbol{R}}{\partial q^j}.   
\end{equation}
Given two tangent vectors to $\mathcal{M}$, e.g., $\boldsymbol {t}_{\mu}=d\boldsymbol{r}_{||}/dq^{\mu}$, we can write the metric on the surface $g_{\mu \nu}\equiv \boldsymbol {t}_{\mu}\cdot\boldsymbol {t}_{\nu}$ and the so-called Weingarten curvature matrix $\alpha_{\mu \nu} \equiv -\boldsymbol {t}_{\mu}\cdot\partial_{\nu}\hat{\boldsymbol{n}}$ to completely characterize the curved surface $\mathcal{M}$, where Greek indices (running over 1 and 2) are used for the adapted coordinates on $\mathcal{M}$. In principle, with the exception of geometric or extrinsic torsion $\beta_{\mu i}^{\ \ k} \eta_{kj} \equiv\hat{\boldsymbol{n}}_i\cdot\partial_\mu \hat{\boldsymbol{n}}_j$, all information of the deformation is intrinsically included in $g_{\mu \nu}$ and $\alpha_{\mu \nu}$, which are also known as the first fundamental form (intrinsic metric) and the second fundamental form (extrinsic curvature), respectively. 
One can then differentiate between extrinsic SDC and intrinsic SDC according to the dependence with respect to the metric on the surface and the Weingarten curvature matrix. However, the electronic Hamiltonian is not directly written in terms of $g_{\mu \nu}$ and $\alpha_{\mu \nu}$, but in terms of the metric tensor $G_{ij}$ and its determinant. Specifically, in the adapted coordinates, the metric tensor and its determinant, i.e.,  
\be
G_{ij} =\left(\begin{array}{ll}
\gamma_{\mu\nu}&
      0 \\
 0  & 1
\end{array}\right)
\ee
%\begin{eqnarray}
%G_{ij}=1/2\left[\mathds{1}(\gamma_{\mu \nu}+1)+\tau_{z}(\gamma_{\mu \nu}-1)\right],
%\end{eqnarray}
and
\begin{equation}
|G|=|\gamma|=|g|(1+q^3 \Tr\alpha_{\mu}^{\ \nu}+(q^3)^2\det\alpha_{\mu}^{\ \nu}),
\end{equation} 
respectively, explicitly involve the intrinsic and extrinsic curvatures through $\gamma_{\mu \nu}=g_{\mu \nu}-2q_{3}\alpha_{\mu \nu}+ q_{3}^{2}\alpha_{\mu\rho}g^{\rho\sigma}\alpha_{\sigma \nu}$ \cite{Costa1981}. Here, we use the convention  $\alpha_{\mu\nu}g^{\nu\beta}=-\alpha_{\mu}^{\ \beta}$ to compute the determinant of $\gamma_{\mu \nu}$, $|\gamma|$. 

%####################################
\subsection{The thin-layer method} In the simplest description of a deformed 2D material, the electronic Hamiltonian  $H_{e}$ includes the kinetic energy $H_{0}=-
\hbar^2\nabla^{2}/2m_{e}$ and a confining potential $V_{c}$ that depends only on the normal coordinate to the surface, which is denoted by $q_3$, i.e., 
\begin{eqnarray}
 H_{e}=H_{0}+V_{c}(q_3).
\end{eqnarray}
In the thin-layer method, instead of assuming that $V_{c}$ becomes infinite outside the surface \cite{Costa1981}, which is unrealistic as it would violate the Heisenberg uncertainty principle, a perturbative expansion should be considered \cite{Maraner95,Jaffe2003,Brandt2015a}. Specifically, the normal coordinate is rescaled, i.e., $q_{3}\rightarrow \epsilon q_{3}$, so the Hamiltonian $H_{e}$ can be expanded in the dimensionless parameter $\epsilon$, which provides an effective Hamiltonian ($\mathcal{H}_{e}$) on the surface based on the perturbative expansion 
\begin{eqnarray}
\epsilon^{2}\mathcal{H}_{e}=\mathcal{H}_{e}^{(0)}+\epsilon\mathcal{H}_{e}^{(1)}+\epsilon^2\mathcal{H}_{e}^{(2)}. 
\end{eqnarray}
Here, $\mathcal{H}_{e}^{(n)}$ is the $n$-th expansion order of the Hamiltonian.
In the ideal scenario, the zero-order (second-order) Hamiltonian $\mathcal{H}_{e}^{(0)}$ ($\mathcal{H}_{e}^{(2)}$) is exclusively dependent on the normal (surface) coordinates. This scenario unravels the electronic structure of deformed two-dimensional materials, freezing the dynamics of the normal direction and decoupling it from the tangential degrees of freedom. 
Consequently, the wavefunction for the electronic Hamiltonian can be written as $\ket{\Psi}=\ket{\varphi}\otimes\ket{\psi}$, where the surface (normal) modes, e.g., $\ket{\psi(q_1,q_2)}$ ($\ket{\varphi(q_3)}$), depend only on the surface (normal) coordinates $q_1$ and $q_2$ ($q_3$). 

The thin-layer method can be summarized in the following steps:
\begin{itemize}
\item[a.] Write the differential operator $H_e$ in the adapted coordinates and rescale it as $\mathcal{H}_e\equiv\left(\frac{|G|^{1/4}}{|g|^{1/4}}\right)H_e\left(\frac{|g|^{1/4}}{|G|^{1/4}}\right)$. In this way, one  define a new wavefunction $\chi$ as $\chi \equiv  \frac{|G|^{1/4}}{|g|^{1/4}}\Psi$.
\item[b.] Insert a dimensionless parameter $\epsilon$ to rescale the normal coordinate, $q^3$, as $q^3 \rightarrow \epsilon q^3$ and expand the rescaled differential operator in a perturbative expansion on $\epsilon$. 
\item[c.] Use a product solution and the normalization condition to decoupling the tangent and the normal degrees of freedom to obtain a effective Hamiltonian in the surface.
\end{itemize}
The dimensional reduction procedure consider here has a structure similar to the Born-Oppenheimer approximation. The first two steps are usually performed separately for each term in the Hamiltonian. 
%The thin-layer method has found extensive application including the coupling of constrained charged particles to electromagnetic fields \cite{Ferrari2008,Ortix2011, Brandt_2017} and Dirac particles \cite{Brandt2016a}. 

%####################################
\subsection{Thin-layer method for the kinetic energy and confinement potential} To better understand the thin-layer method and its subsequent implementation for the SDC study, we illustrate it for the kinetic energy and the confinement potential for spinless electrons as an example. We now turn our attention to the use of the metric tensor $G_{ij}$ and its determinant $|G|$ to redefine the electronic Hamiltonian $H_{e}$ in $\mathcal{M}$ (step a). Our starting point is the electronic Hamiltonian  $H_{e}$ including the kinetic energy $H_{0}=-
\hbar^2\nabla^{2}/2m_{e}$ and a confining potential $V_{c}$.

\subsubsection{Confinement potential} For step a, one can note that the potential $V_{c}$ has been assumed to depends only on the normal coordinate $q_{3}$ and hence, there is no change in it mathematical form. For step b, the procedure is simplified considering that, due to the deep minima, the confinement potential can be approximated, before rescale, as $V_c=\bar{V}_{c}+O(q_{3}^{3})$ \cite{Maraner95}. Here, the potential $\bar{V}_{c}$ is symmetric with respect to the surface, e.g., the harmonic confinement potential $\bar{V}_{c}=\frac{1}{2\epsilon^4}m\omega^2 (q_3)^2$. 
%meanwhile $O(q_{3}^{3})$ includes contributions to the potential that depend on powers of $q_3$ higher than $q_3^{2}$. 
After rescaling, we can expand $\mathcal{V}_c$ in an arbitrary series in $\epsilon$ (step b),
\begin{eqnarray}\nonumber
\epsilon^{2}\mathcal{V}_{c}&=&\mathcal{V}_{c}^{(0)}+\epsilon \mathcal{V}_{c}^{(1)}+\epsilon^{2} \mathcal{V}_{c}^{(2)}+O(\epsilon^{3})\\
&=&\frac{1}{2}m\omega^2 (q_3)^2+O(\epsilon^{3}),
\end{eqnarray}
one can note that the order zero $\mathcal{V}_{c}^{(0)}$ is actually the quadratic term, while $\mathcal{V}^{(1)}_{c}$ and $\mathcal{V}^{(2)}_{c}=0$. We would like to emphasize that by imposing that $O(q_{3}^{3})$ in $V_c$ is independent of $\epsilon$, the Hamiltonian describing the surface, $\mathcal{H}^{(1)}_e$, becomes entirely independent of the explicit form of $V_c$ \cite{Maraner95}. Indeed, the validity of the thin-layer method relies on the fact that the energy scale of $V ^{(0)}$ should be much larger than that of $\epsilon \mathcal{H}^{(1)}_e$ and subsequent terms.

%{\color{red}A realistic approach for the potential could be its association with the periodic effective potential generated by the electrostatic interaction between the electron and the $P$ nuclei in a two-dimensional periodic solid, i.e., $V_{eff}(\boldsymbol{r})=-\left(e^{2}/2\right)\sum\limits_{a}^P Z_{a}/|\boldsymbol{r}_{a}|$ with $\boldsymbol{r}_{a}=\boldsymbol{r}-\boldsymbol{R}_{a}$ being the electron position with respect to the $c$-nth nuclei with atomic number $Z_{a}$ at the positions $\boldsymbol{R}_{a}$. This approach can potentially change the direction of the dipole when the nanomaterial is deformed.} 

\subsubsection{Kinetic energy}
On the curved surface $\mathcal{M}$, the kinetic energy $H_{0}=-
\hbar^2\nabla^{2}/2m_{e}$ is rewritten as (step a)
\begin{equation}\label{kineticterm3D}
H_{0}=-\frac{\hbar^2}{2m_{e}|G|^{1/2}}\partial_i |G|^{1/2}G^{ij}\partial_j.
\end{equation}
Here, $|G|=|g|(1+q_3 \Tr\alpha_{\mu}^{\ \nu}+q_{3}^{2}\det\alpha_{\mu}^{\ \nu})$ is the determinant of the metric tensor in the adapted curved coordinates, and $m_{e}$ is the electron mass. Rescaling and expanding the kinetic energy powers of $\epsilon $ (step a and b)
\begin{eqnarray}
\epsilon^{2}\mathcal{H}_{0}=\mathcal{H}_{e}^{(0)}+\epsilon\mathcal{H}_{0}^{(1)}+\epsilon^2\mathcal{H}_{0}^{(2)}, 
\end{eqnarray}
we find that 
\begin{equation}
\mathcal{H}_{0}^{(0)}=-\frac{\hbar^2}{2m}\partial_{q_3}^2+\mathcal{V}_{c},
\end{equation} 
\begin{equation}
\mathcal{H}_{0}^{(1)}=0,
\end{equation}  
and
\begin{equation}
\mathcal{H}_{0}^{(2)}=\left[-\frac{\hbar^2}{2m|g|^{1/2}}
\partial_{\mu}g^{\mu\nu}|g|^{1/2}\partial_{\nu}+\mathcal{V}_g\right].
\end{equation}  

The geometrical deformation effects emerge only on the second order of expansion as a quantum geometric potential, i.e., 
$\mathcal{V}_{g}=-\frac{\hbar^2}{2m}(\text{H}^2-\text{K})$ and a metric tensor field $g^{\mu \nu}$ in the curved surface~\cite{Jesen1971,Costa1981,Maraner95,Jaffe2003}.
As previously mentioned, two contributions can be identified: the mean curvature $\text{H}\equiv-\Tr \alpha_{\mu}^{\nu}/2$ and the Gaussian curvature $\text{K}\equiv\det \alpha_{\mu}^{\nu}$. The Gaussian curvature, which is half of the scalar curvature, measures the intrinsic curvature of a surface, independent of its embedding space, while the mean curvature depends on its extrinsic geometry. Notably, empirical evidence of the geometric potential within quantum systems has been compellingly demonstrated in experiments using the thin-layer method, such as in metamaterials that exhibit a photonic topological crystal \cite{Szameit2010} and in the Riemannian geometric effects observed within a one-dimensional metallic 
$C_{60}$ polymer \cite{Onoe2012}.

On the other hand, note that $\mathcal{H}_{0}^{(0)}$ and $\mathcal{H}_{0}^{(2)}$ depends only on the normal and surface coordinates, respectively. Up to the second-order in $\epsilon$, there is no coupling between $q^3$ and $q^{\mu}$, which justifies the separation of variables $\ket{\Psi}=\ket{\varphi(q^3)}\otimes\ket{\psi(q^1,q^2)}$, where $\ket{\varphi(q^3)}$ is the ground state of $\mathcal{H}^{(0)}$. Indeed, in the kinetic energy and confinement potential, the first order terms in the expansion are null. The presence of the term $\mathcal{H}_{e}^{(1)}$ could result in the coupling of normal and surface modes.

In this illustrative demonstration of the formalism, as very well established for spinless electrons~\cite{Teufel2010}, we verify that there is no coupling between the surface and normal modes, i.e., the surface modes $\ket{\psi(q_1,q_2)}$ and normal modes $\ket{\varphi(q_3)}$ are only described by the Hamiltonian at second and zero order in $\epsilon$, respectively. The effective Hamiltonian for the surface $\mathcal{H}_{e}^{(2)}$ is nothing than the term $\mathcal{H}_{0}^{(2)}$. Note that when $\epsilon$ is small (deep confining potential), the confining energy ($\mathcal{H}_{e}^{(0)}$) is significantly larger than the kinetic energy ($\epsilon^2 \mathcal{H}_{0}^{(2)}$); thus, the transitions between the normal and tangential modes are suppressed \cite{Teufel2010}. The established steps and analysis of the physical meaning of the terms in the expansion are the starting point to include the SOC in the surface Hamiltonian and the posterior study of the SDC.

\section{Spin-deformation coupling based on the phenomenological linear Rashba effect} 
In this section, we use the previous parametrization of the surface deformation and the thin layer method for an electronic Hamiltonian including the SOC in polar compounds, which is introduced throughout the Rashba effect, i.e., 
\begin{equation}
H_{e}=H_{0}+V_{c}+H_{R}.
\end{equation} 
As demonstrated below, when this procedure is performed, in contrast to the kinetic energy, the SOC introduces first order in $\epsilon$ terms, i.e., $\mathcal{H}_{R}^{(1)}\neq 0$, coupling normal and surface modes. 
Consequently, deriving the effective dynamics in 2D translates into the challenge of formulating a strategy to eliminate the coupling mediated by $\mathcal{H}_{R}^{(1)}$. Despite the anticipated coupling between normal and surface modes, we follow the thin-layer method for the Rashba effect, analogous to Refs. \cite{Chang2013a,Wang2017a,Taichi2011,Siu2018}, in order to provide an understanding of the relationship between both the Rashba effect and surface deformation with the SDC, before introducing an $SU(2)$ non-Abelian gauge field for the Rashba spin-orbit coupling.
%Although we have anticipated that the thin-layer method for the Rashba effect procedure conduces to a coupled Hamiltonian, it provides an understanding of the relation between both the Rashba effect and surface deformation with the SDC, prior to introducing an $SU(2)$ non-Abelian gauge field for the Rashba spin-orbit coupling.} 

The SOC is a relativistic correction that results from the interaction between the magnetic moment associated with the spin of electrons and the magnetic field generated by the electron's motion through a potential gradient $\boldsymbol{\nabla}\phi$, i.e., $H_{\text{SOC}}=-\frac{e\hbar}{4m^{2}c^{2}}\boldsymbol{\sigma}\cdot \left(\boldsymbol{\nabla}\phi\times \boldsymbol{p}\right)$. Emmanuel Rashba noticed that this term for an electron with mass $m$, momentum $\boldsymbol{p}$ and spin $\boldsymbol{s}=\frac{\hbar}{2}\boldsymbol{\sigma}$ can also be induced by an external electric field \cite{bychkov1984properties}. Consequently, when the potential gradient is induced by a uniform electric field perpendicular to the surface, the SOC can be rewritten as $\mathcal{H}_{R}=\lambda_{R}(\sigma_{x}k_{y}-\sigma_{y}k_{x})\hat{z}$, which is typically referred to as the phenomenological linear Rashba effect. In the last decade, it was realized that the electric dipole in polar compounds induces a potential gradient leading to an intrinsic Rashba effect \cite{Ishizaka2011-bo}, referred to as bulk Rashba effect \cite{MERAACOSTA2020145}. A spin splitting in the wavevector $\boldsymbol{k}$ space as well as a helical spin polarization are the main characteristics of the Rashba effect \cite{PhysRevB.94.041302}, which have been at the heart of the potential application in spintronics \cite{Manchon2015-lp}. This interaction is particularly significant in atoms where the relativistic effects become appreciable, such as those with high atomic numbers. The magnitude of $\lambda_{R}$ varies across materials and is influenced by the atomic number and the electronic environment \cite{MERAACOSTA2020145,PhysRevB.104.104408}. Even if the electric dipole is arbitrarily fixed along the normal direction, a spin-dependent geometrical potential could emerge \cite{Bihlmayer2022}. 

On the curved surface $\mathcal{M}$, the phenomenological Rashba effect
$H_{R}=S_{ij}\tilde{\sigma}^i p^j$ is rewritten as 
\begin{equation}
H_{R}=-i \hbar \mathcal{S}_j G^{jk}\partial_k, 
\end{equation} 
where we define $\mathcal{S}_{j}=S_{ij}\tilde{\sigma}^i$. Here, $\tilde{\sigma}^i$ are the space-dependent Pauli matrices, and $S_{ij}=0$ if $i=j$. 
The space-dependent $\sigma$-matrices and the Rashba tensor $S_{ij}$
can be defined as follows 
\begin{eqnarray}
\tilde{\sigma}^{i} &=&\frac{\partial q^{i}}{\partial x^{s}} \sigma^{s}, \\
S_{i j} &=&\frac{\partial x^{s}}{\partial q^{i}} \frac{\partial x^{t}}{\partial q^{j}} S_{s t}, 
\end{eqnarray}
where $\sigma^{s}$ and $S_{s t}$ are the associated Pauli matrices and the Rashba tensor in the Cartesian coordinate system, respectively \cite{Chang2013a,Wang2017a}. Following the confining potential approach, we first rescaled the Rashba Hamiltonian as
\begin{eqnarray}
\mathcal{H}_{R}&\equiv&\frac{|G|^{1/4}}{|g|^{1/4}}H_{R}\frac{|g|^{1/4}}{|G|^{1/4}}.
\end{eqnarray}
For a consistent perturbative expansion, it is crucial to take into account that all functions of $q_3$ and the derivatives with respect to $q_3$ have to be rescaled by $q_3 \rightarrow \epsilon q_3$, thus
\be
\frac{\partial}{\partial q_3}\rightarrow \frac{1}{\epsilon}\frac{\partial}{\partial q_3},
\ee
and $\mathcal{S}_{j}(q^{\mu},\epsilon q_3)$ now depends on $\epsilon$, therefore it can be Taylor-expanded as follows 
\be
\mathcal{S}_{j}(q^{\mu},\epsilon q_3)=
\mathcal{S}_{j}(q^{\mu},0)+\epsilon q_3\left. \partial_3 \mathcal{S}_{j}(q^{\mu},q_3)\right|_{q_3=0}+\ldots . \ \ \
\ee
Now considering the Rashba effect by expanding the Hamiltonian $\mathcal{H}_{R}$ up to second order in $\epsilon$, i.e., $\epsilon^2\mathcal{H}_{R}=\mathcal{H}_{R}^{(0)}+\epsilon\mathcal{H}_{R}^{(1)}+\epsilon^2\mathcal{H}_{R}^{(2)}+\ldots$, we obtain 
\begin{equation}
\mathcal{H}_{R}^{(0)}=0
\end{equation}
\begin{equation}
\mathcal{H}_{R}^{(1)}=-i\hbar\mathcal{S}_{3}\partial^{3},
\end{equation}
and
\begin{equation}
\mathcal{H}_{R}^{(2)}=-i \hbar\left(\bar{\mathcal{S}}_{\mu} g^{\mu\nu}\partial_{\nu}+\bar{\mathcal{S}}_{3}\text{H}+ \frac{\partial \bar{\mathcal{S}}_{3}}{\partial q^3}q^3\partial^{3}
\right),
\label{FinalEffectiveRashba0}
\end{equation}
where we use the following notation $\bar{\mathcal{S}}_{j}=\mathcal{S}_{j}(q^{\mu},0)$ and $\partial^3 \bar{\mathcal{S}}_{3}=\left.\partial^3 \mathcal{S}_{3}(q^{\mu},q^{3})\right|_{q^{3}=0}$. 

We find that there is no contribution from the surface Rashba effect to the zero-order perturbative expansion (i.e., $\mathcal{H}_{R}^{(0)}=0$). In $\mathcal{H}_{R}^{(2)}$, the first term resembles that expected from the Rashba effect on a curved surface, the intrinsic SDC. The last two terms present in $\mathcal{H}_{R}^{(2)}$ can be interpreted as the extrinsic SDC, i.e., the coupling of the mean curvature with the spin and another term associated with $q^3\partial^{3}$, which should be adequately frozen in the ground state of the normal modes. Moreover, we note that $\mathcal{H}_{R}^{(1)}$ couples the surface and normal modes, indicating the Rashba effect forbids writing the wavefunction as $\ket{\Psi}=\ket{\varphi}\otimes\ket{\psi}$. 

As implied by the perturbative expansion, effects intrinsic to the surface electronic dynamics should not modify the normal modes. Therefore, the transitions between the normal modes promoted by the SOC term $\mathcal{H}_{R}^{(1)}$ should be carefully studied, as the term $\bar{\mathcal{S}}_{3}\text{H}$ could be a fictitious collateral consequence of a methodologically inconsistent approach. 
%the first term is the Rashba effect in the curved surface, the second term represents spin-deformation coupling (i.e., the combining effect of the Rashba SOC and the extrinsic curvature), and the last term is an \textcolor{red}{energy shift} introduced for a given out-of-plane state. 
%, i.e., the out of plane modes are frozen with respect to the in-plane electronic dynamics. 
%\textcolor{red}{Additionally, if $\epsilon$ is small with respect to kinetic energy normal to the surface (deep confining potential), the confining energy ($\mathcal{H}_{0}^{(0)}+\tilde{V}_{c}$) is much larger than the in-plane kinetic energy ($\mathcal{H}_{\text{K}}^{(2)}$), thus transitions between normal and in-plane modes are energetically forbidden.}
This is commonly achieved by imposing $\bra{...} \partial_{q_3} f(q_{3})\ket{...}=0$~\cite{Chang2013a,Wang2017a}, which could result in the inclusion of fictitious terms in the SDC. 
Indeed, one could think that the control of the spin can be achieved throughout geometrical effects, which is erroneous, as we demonstrate below using a non-Abelian field to describe the Rashba SOC. %{\color{blue}As we will demonstrate below, there is only one spin-independent potential that represents the sole contribution of the extrinsic curvature to the system's dynamics, which implies the absence of a extrinsic SDC.}

%As we will demonstrate, this spin-independent potential constitutes the exclusive contribution of the extrinsic curvature to the dynamics of the system, which implies the absence of a extrinsic SDC.

%%%%%%%%%%%%%%%%%%%%%%%%%%%%%%%%%
\section{Spin-deformation coupling based on a non-Abelian gauge field}
%%%%%%%%%%%%%%%%%%%%%%%%%%%%%%%%%
In this section, inspired by a common approach in strongly correlated many-body physics that utilizes a unitary gauge transformation to disentangle certain degrees of freedom
~\cite{PhysRevLett.126.153603}, we search for an appropriate unitary transformation to disentangle surface and normal quantum dynamics. This approach yields an expansion (step b in the thin-layer method) with a null first-order term, i.e., $\mathcal{H}_{R}^{(1)}=0$. The general idea is that a highly entangled quantum state in the original frame can be expressed as a factorable state after the transformation, which has been used for analyzing quantum impurity systems~\cite{PhysRev.90.297,Frohlich1952-ei,Silbey1984-pf,Silbey1984-pf,PhysRevLett.121.026805}, constructing low-energy effective models~\cite{PhysRev.149.491,Wegner1994-ah,PhysRevD.48.5863,Bravyi2011-hr}, and solving many-body localization or electron-phonon problems~\cite{Imbrie2016-ep,PhysRevLett.125.180602}. 
The starting point in searching for a convenient gauge transformation is to interpret the Rashba SOC as an effective non-Abelian vector potential—a vector field associated with electron spin \cite{Frohlich1993,Jin_2006,Tokatly2008}, where different directions of spin can result in different gauge fields associated with Rashba SOC.
%A key characteristic of non-Abelian gauge fields is that they can take on multiple values, unlike Abelian gauge fields. 
%%%%%%%%%%%%%%%%%%%%%%%
\subsubsection{The Rashba effect as a non-Abelian gauge field}
%%%%%%%%%%%%%%%%%%%%%%%
In order to maintain generality, we include external semiclassical electromagnetic fields, encompassing a broad class of systems, among which the Zeeman coupling is covered. In this scenario, the electronic Hamiltonian is 
\begin{equation}
H_{e}=H_{0}+H_{R}+H_{Z}+H_{P},
\end{equation} 
which includes the kinetic energy 
\begin{equation}
H_{0}=\frac{\left(\boldsymbol{p}-e\boldsymbol{\mathcal{A}}\right)^2}{2 m},
\end{equation}
the Rashba effect 
\begin{equation}
H_{R}=\frac{e \hbar}{4 m^2 c^2} \left(\boldsymbol{p}-e\boldsymbol{\mathcal{A}}\right) \cdot(\boldsymbol{\sigma} \times \nabla V),
\end{equation}
the Zeeman effect
\begin{equation}
H_{Z}=g_{2D}\frac{e\hbar}{2m}\boldsymbol{\sigma} \cdot \boldsymbol{B},
\end{equation}
and the electric potential $H_{P}=e\mathcal{A}_0$. 
Electrons in semiconductors can have an effective $g$-factor that differs substantially from the free-electron value $g_0=-2$ as a consequence of the SOC, which couples the orbital motion with the spin degree of freedom~\cite{PhysRev.114.90}. This is introduced in the Zeeman effect through a $g$-factor for 2D materials, i.e., $g_{2D}$, which can even differ between conduction and valence bands~\cite{PhysRevLett.126.067403}.  Here, $\boldsymbol{\sigma}$ represents Pauli spin matrices, $\boldsymbol{p}$ is the momentum operator, $\boldsymbol{\mathcal{A}}$ and $A_0$ are the vector potential and the scalar potential of the electromagnetic field, respectively, and $V$ is the crystal potential. As it is well established \cite{Frohlich1993}, a $SU(2)$ non-Abelian  gauge field $\boldsymbol{\mathcal{W}}$ captures the spin-orbit interaction and the Zeeman coupling when the following identifications are made:
\begin{equation}
\mathcal{W}_i=\frac{e\hbar}{m c^2} \varepsilon_{i j a} E_j \tau^a, \ \ \ \ \  \mathcal{W}_0=-\frac{e\hbar}{m}B_a\tau^a,
\end{equation}
with $\tau^a=\sigma^a / 2$. 
For instance, if the electric dipole is perpendicular to the two-dimensional material, i.e., $E_{x}=E_{y}=0$, then the non-Abelian gauge field is 
\begin{equation}
\mathcal{W}_x=-\frac{2m\alpha}{\hbar} \tau^y, \ \ \ \ \mathcal{W}_y=\frac{2m\alpha}{\hbar} \tau^x, \ \ \ \ \mathcal{W}_0=-\frac{e\hbar}{m}B_a\tau^a.     
\end{equation}
This $SU(2)$ non-Abelian  gauge field  satisfies the Coulomb gauge restriction, $\nabla \cdot \boldsymbol{\mathcal{W}}=0$ \cite{Shikakhwa_2012}, and enables the rewriting of the electronic Hamiltonian as
\begin{equation}
H_{e}=\frac{(\boldsymbol{p}-e\boldsymbol{\mathcal{A}}-\boldsymbol{\mathcal{W}})^2}{2 m}-\frac{1}{2 m} \boldsymbol{\mathcal{W}} \cdot \boldsymbol{\mathcal{W}}+\mathcal{W}_0+e\mathcal{A}_0.
\label{NonDecoupledH}
\end{equation}
When $\mathcal{W}_{0}=0$ and $\mathcal{A}=0$, the electronic Hamiltonian is reduced to the Rashba effect, as a matter of discussion we preserved all terms in the Hamiltonian. 

Before decoupling the surface and normal modes, we first introduce a gauge covariant derivative $\mathcal{D}_{i}=\partial_{i}-\frac{ie}{\hbar}\mathcal{A}_{i}-\frac{i}{\hbar}\mathcal{W}_{i}$
and the generalized gauge field $\boldsymbol{\mathcal{Z}}=e\boldsymbol{\mathcal{A}}-\boldsymbol{\mathcal{W}}$. In the adapted curved coordinates, Eq. (\ref{NonDecoupledH}) takes the form
\begin{eqnarray}\nonumber\label{HamiltoniCurviGauge}
 H_{e}&=&-\frac{\hbar^2}{2m|G|^{1/2}}\mathcal{D}_i
|G|^{1/2}G^{ij}\mathcal{D}_j-\frac{1}{2m}G^{ij}\mathcal{W}_{i}\mathcal{W}_{j}\\
&&+\mathcal{W}_0+e\mathcal{A}_0+V_c.
\end{eqnarray}
%Here, $\mathcal{Z}(q^{\mu},\epsilon q^i)$ and $\mathcal{W}(q^{\mu},\epsilon q^i)$ also depend on $\epsilon$ and hence, a Taylor expansion should also be consider for generalized gauge field, e.g., 
%\begin{equation}
%$\mathcal{Z}(q^{\mu},\epsilon q^i)\approx\bar{\mathcal{Z}}_{\mu}+\epsilon q^j\left.\frac{\partial \mathcal{Z}(q^{\mu},q^i)}{\partial q^j}\right|_{q=0},
%\end{equation}
%where $\bar{\mathcal{Z}}_{\mu}=\mathcal{Z}(q^{\mu},0)$. 
%The expansion on $\epsilon$ for $\mathcal{W}(q^{\mu},\epsilon q^i)$  is completely analogous, with $\bar{\mathcal{W}}_{\mu}$ being equal to $\mathcal{W}(q^{\mu},0)$).
The Hamiltonian above is gauge-invariant, except for the term $G^{ij}\mathcal{W}_{i}\mathcal{W}_{j}/2m$, which breaks the SU(2) gauge invariance. This term is commonly absorbed into a scalar potential $\phi(x)$. 
If the expansion (step b of the thin-layer method) is directly applied to this electronic Hamiltonian (Eq. (29)), the result obtained is analogous to the previous section. Specifically, the coupling between normal and surface modes arises from the first-order term of the expansion. However, in the non-Abelian formulation, $\mathcal{H}_{e}^{(1)}$ is proportional to the product of the field $\mathcal{W}_i$ and the derivative $\partial^{i}$. As demonstrated below, a suitable unitary transformation, $\mathbf{U}$, is introduced to address this term.

%, concealing the lack of invariance in the Hamiltonian \cite{Tokatly2008}. However, this term must be explicitly considered in our analysis, since it introduces contributions to the effective Hamiltonian.
%In this description, when the thin-layer is implemented, as demonstrated in detail in the supplementary material, and consistent with the previously explored procedure, the surface and normal modes are coupled via the first-order expansion term $\mathcal{H}_{e}^{(1)}$, i.e., $\mathcal{H}^{(1)}_{e}=-\frac{i}{m}\hbar\bar{\mathcal{Z}}_{3}\partial^{3}$ with $\bar{\mathcal{Z}}_{3}=\mathcal{Z}_3(q^{3},0)$. Among the strategies previously implemented to eliminate $\mathcal{H}_{e}^{(1)}$ is the condition $\bra{...} \partial^3 f(q{3})\ket{...}=0$ \cite{Chang2013a,Wang2017a}. However, this proves inconsistent with the approximation, as the energy scales of $\mathcal{H}_{e}^{(1)}$ are larger than those of $\mathcal{H}_{e}^{(2)}$. As we will demonstrate, this inconsistency results in the exclusion of relevant terms and the inclusion of fictitious ones.

\subsubsection{Unitary gauge transforamtion}

In this subsection, we aim to decouple the fields in a more general context than previously discussed. We consider a particle confined to an $m$-dimensional submanifold ($\mathcal{M}$) within an $n$-dimensional manifold ($\mathcal{N}$). While we have so far focused on the case of a curved surface with $m=2$ and $n=3$, this generalization allows us to extrapolate our results to other scenarios, such as a curved wire with $m=1$ and $n=3$ (see Appendix \ref{AppenAmanifold} for a detailed explanation of the notation).

Analogous to the Abelian case  \cite{Ferrari2008,Ortix2011,Brandt2015a,Brandt_2017}, we can decouple the dynamics by performing a suitable gauge transformation, 
\begin{equation}\label{nonAbeliangaugetransform}
\mathbf{U}= e^{-i T(x^{\mu},\epsilon q^i)/\hbar }    
\end{equation}
with 
\begin{equation}
T(x^{\mu},\epsilon q^i)=\epsilon \bar{\mathcal{W}}_iq^i+\frac{\epsilon^2}{2}\frac{\partial  \bar{\mathcal{W}}_i}{\partial q^j}q^jq^i+\mathcal{O}(\epsilon^2 ).
\end{equation}
Here, the notation $\bar{\boldsymbol{\mathcal{W}}}=\boldsymbol{\mathcal{W}}(q^{\mu},0)$ has been introduced. The transformation $\mathbf{U}$ has the general form used to disentangle strong correlation in many-body systems ($\mathbf{U}\equiv \exp\left(-i\epsilon T/\hbar \right)$ with a proper choice of $T$). 
Note that this transformation can always be performed, even if the Hamiltonian (\ref{HamiltoniCurviGauge}) is not gauge invariant. 

The finite gauge transformation for the component $\mathcal{W}_i$ of non-Abelian field is
\begin{equation}
\mathcal{W}'_{j}=\mathbf{U}\mathcal{W}_{j}\mathbf{U}^{-1}-\frac{i\hbar}{\epsilon}(\partial_{j} \mathbf{U})\mathbf{U}^{-1}.
\end{equation}
For the second term in this gauge transformation, we obtain  
\begin{equation}
\frac{i\hbar}{\epsilon}(\partial_{j} \mathbf{U})\mathbf{U}^{-1}=\bar{\mathcal{W}}_j+\frac{\epsilon}{2}\frac{\partial  \bar{\mathcal{W}}_j}{\partial q^k}q^k+\frac{\epsilon}{2}\frac{\partial  \bar{\mathcal{W}}_k}{\partial q^j}q^k+\mathcal{O}(\epsilon^2 ).
\end{equation}
To calculate the first term in this non-Abelian gauge transformation, it is necessary to expand $\mathbf{U}$ on $\epsilon$, i.e,  
\begin{equation}
\mathbf{U}= \sum_{k=0}^{\infty} \frac{1}{k!}\left(i\epsilon T(x^{\mu}, \epsilon q^i)\right)^k,  
\end{equation}
which results in 
\begin{equation}
\mathbf{U} \mathcal{W}_j \mathbf{U}^{-1} = \bar{\mathcal{W}}_j + \epsilon \frac{\partial \bar{\mathcal{W}}_j}{\partial q^i} q^i + i \epsilon q^i [\bar{\mathcal{W}}_i, \bar{\mathcal{W}}_j] + \mathcal{O}(\epsilon^2)
\end{equation}
where, we have used that $\mathbf{U}\mathcal{W}_{j}\mathbf{U}^{-1}= \mathcal{W}_{j}\mathbf{U}\mathbf{U}^{-1}+ [\mathbf{U},\mathcal{W}_{j}]\mathbf{U}^{-1}$.
Therefore, by applying a gauge transformation to the normal coordinates, we obtain
\begin{equation}
\mathcal{W}'_{i}=-\epsilon \frac{q^j}{2}\bar{\mathcal{G}}_{ij}+\mathcal{O}(\epsilon^2 ),
\end{equation}
where $\bar{\mathcal{G}}_{ij}=\left.\mathcal{G}_{ij}\right|_{q=0}$
is the non-Abelian $SU(2)$ field tensor of the normal components to the surface \cite{Liang2020a}. There is only one normal component on a surface, so the field tensor is identically zero, since it is antisymmetric. Consequently, through the application of a gauge transformation, it becomes feasible to eliminate, to second order in $\epsilon$, the field component $\mathcal{W}_{q^3}$ from all gauge invariant terms. When the co-dimension ($n-m$) is greater than 1, contributions related to the field tensor will appear, see Appendix \ref{AppendDecou} for more details.

By applying a gauge transformation to the coordinates on the surface, one can show that $\mathcal{W}'_{\mu}=\bar{\mathcal{W}}_{\mu}+\mathcal{O}(\epsilon)$. Here, only terms of order \(\epsilon\) are considered, as they are the only ones that contribute to the effective Hamiltonian $\mathcal{H}_e^{(2)}$, see Eq. (\ref{effectivebeforegt}). 

On the other hand, for the non-gauge invariant term in the Hamiltonian (Eq. (\ref{HamiltoniCurviGauge})), we note however that this term can be written
as 
\begin{eqnarray}\nonumber     \mathcal{W}_{i}\mathcal{W}^{i}+\mathcal{W}_{\mu}g^{\mu \nu} \mathcal{W}_{\nu} &=&  \mathcal{W}_{ia}\tau^a   \mathcal{W}^{i}_{b}\tau^b + \mathcal{W}_{\mu a}\tau^a   \mathcal{W}^{\mu}_{b}\tau^b
\\ 
&=& (\mathcal{W}_{ia}\mathcal{W}^{ia} +\mathcal{W}_{\mu a}\mathcal{W}^{\mu a}) \mathbf{I} _{2},
\end{eqnarray}
where we have used the properties of matrices $\tau^a$ and $\mathbf{I}_{2}$ is the $2\times 2$ identity matrix, thus
\begin{equation}\label{noninvariantbefore}
\mathbf{U}\left[\mathcal{W}_{i}\mathcal{W}^{i}+\mathcal{W}_{\mu}g^{\mu \nu} \mathcal{W}_{\nu}\right]\mathbf{U}^{-1}=\mathcal{W}_{i}\mathcal{W}^{i}+\mathcal{W}_{\mu}g^{\mu \nu} \mathcal{W}_{\nu}.
\end{equation}
Hence, this term remains unaffected by the this unitary transformation. 

Note that in Eq. (\ref{noninvariantbefore}), the fields are those before performing the gauge transformation. For the surface components, there is no difference because $\bar{\mathcal{W}}'_{\mu}=\bar{\mathcal{W}}_{\mu}$. However, for normal components, we use the notation $\mathcal{W}_i=\mathcal{W}^B_i$ where the superscript indicates that the field is included before the gauge transformation.

Once this transformation is performed, as established in the thin-layer method, the normal coordinates is rescaled, i.e., $q \rightarrow \epsilon q$, and the perturbative expansion for $\mathcal{H}_{e}$ in powers of $\epsilon$ is considered. 
%%%%%%%%%%%%%%%%%%%%%%%%%
\subsubsection{Surface Hamiltonian for the non-Abelian gauge field}
%%%%%%%%%%%%%%%%%%%%%%%%%
When the normal coordinates are rescaled $q \rightarrow \epsilon q$ and the electronic Hamiltonian is expanded in powers of $\epsilon$, i.e, $\epsilon^{2}\mathcal{H}_{e}=\mathcal{H}_{e}^{(0)}+\epsilon\mathcal{H}_{e}^{(1)}+\epsilon^2\mathcal{H}_{e}^{(2)}$, we have to consider that $\boldsymbol{\mathcal{Z}}(q^{\mu},\epsilon q^i)$ and $\boldsymbol{\mathcal{W}}(q^{\mu},\epsilon q^i)$ are also dependent on $\epsilon$. Specifically, by introducing the Taylor expansions 
%we expand $A_B$ in a Taylor series 
\begin{equation}\label{boldsymbolz}
\boldsymbol{\mathcal{Z}}(q^{\mu},\epsilon q^i)=
\boldsymbol{\mathcal{Z}}(q^{\mu},0)+\epsilon q^j\left.\frac{\partial \boldsymbol{\mathcal{Z}}(q^{\mu},q^i)}{\partial q^j}\right|_{q=0}+\ldots 
\end{equation}
and 
\begin{equation}\label{boldsymbolw}
\boldsymbol{\mathcal{W}}(q^{\mu},\epsilon q^i)=
\boldsymbol{\mathcal{W}}(q^{\mu},0)+\epsilon q^j\left.\frac{\partial \boldsymbol{\mathcal{W}}(q^{\mu},q^i)}{\partial q^j}\right|_{q=0}+\ldots, \end{equation}
we can obtain the electronic Hamiltonian up to the second order in $\epsilon$ (see Appendix for more details). After applying the Abelian (\ref{Abeliangaugetransform}) and non-Abelian (\ref{nonAbeliangaugetransform}) gauge transformations, we can decouple the normal and surface modes, as evidenced by a null first-order term, and zero (second) order depending only on the normal (surface) coordinate, i.e.,
\begin{equation}
\mathcal{H}^{(0)}_{e}=-\frac{\hbar^2}{2m}\partial_3\partial^3 +\mathcal{V}_c,
\end{equation}
\begin{equation}
\mathcal{H}^{(1)}_{e}=0,
\end{equation}
and
\begin{equation}
\mathcal{H}^{(2)}_{e}=\frac{-\hbar^2}{2m|g|^{1 / 2}}\mathcal{D}_\mu g^{\mu \nu}|g|^{1/2}\mathcal{D}_\nu-\frac{\bar{\mathcal{W}}_{\mu}g^{\mu \nu} \bar{\mathcal{W}}_{\nu}}{2m}+\mathcal{Z}_0+\mathcal{V}_g-{\mathcal{\phi}},
\label{FinalEffectiveRashba1}
\end{equation}
where we have introduced the scalar potential   
\begin{equation} 
\mathcal{\phi}=\frac{1}{2m}\bar{\mathcal{W}}^B_{3}\bar{\mathcal{W}}^{B3},  \end{equation}
where the covariant derivative is now expressed as $\mathcal{D}_{\mu}=\partial_{\mu}-\frac{ie}{\hbar}\bar{\mathcal{A}}_{\mu}-\frac{i}{\hbar}\bar{\mathcal{W}}_{\mu}$. The effective surface electronic Hamiltonian $\mathcal{H}^{(2)}_{0}$ is decoupled from normal states. From the perspective of the SDC, the physical meaning of Eq. \ref{FinalEffectiveRashba1} is completely different from the surface Hamiltonian obtained from the phenomenological Rashba effect (Eq. (\ref{FinalEffectiveRashba0})), specifically: \textit{i}) for the intrinsic SDC, the Rashba effect on a curved surface is incorporated through the surface non-Abelian gauge field $\bar{\mathcal{W}}_{\nu}$, which couples with the metric and consequently with the intrinsic curvature and \textit{ii}) the extrinsic SDC is absent in the Hamiltonian including the term $\bar{\mathcal{S}}_{3}\text{H}$, as we anticipate by analogy with the interaction between electromagnetic fields and geometrical deformations. This can be physically interpreted by considering that the gauge invariance of the Hamiltonian $\tilde{\mathcal{H}}_{e}=\mathcal{H}_{e}+\frac{1}{2 m} \boldsymbol{\mathcal{W}} \cdot \boldsymbol{\mathcal{W}}$ forbids the extrinsic SDC term $\bar{\mathcal{S}}_{3}\text{H}$. 
Note that although the second term in the electronic Hamiltonian Eq.(\ref{NonDecoupledH}) is not gauge invariant, it does not play a role in the interaction that mediates the coupling between the surface and normal modes. 

Our results show that in opposite to the prevailing consensus, the geometric deformations of the extrinsic curvature have no effect on the spin dynamics and cannot be used to control the spin degree of freedom. 
The use of non-Abelian gauge fields to address SOC in the context of curved quantum wires is of particular significance \cite{Liang2020a}. Curiously, the coupling term $\bar{\mathcal{W}}_{3}\text{H}$ does not appear in the effective Hamiltonian in \cite{Liang2020a}. The absence of extrinsic SDC, which contradicts several previous works \cite{Zhang2007a,Liu2011a,Gentile2013a,Ortix2015ab,Gentile2015a,Ying2016a,Pandey2018a,Salamone2022a}, seems to have been overlooked. 
Naturally, the consequences arising from the fictitious term $\text{H}\tilde{\sigma}^i$ must be reevaluated. 
For instance, although it has been suggested that in the presence of non-zero curvature, the electron spin acquires a finite out-of-plane binormal component\cite{Ying2016a,Gentile2022}, our results indicate that the spin textures remain confined to a plane. This suggests that earlier interpretations based on the existence of out-of-plane spin texture contributions may also need to be reconsidered. 
For instance, in curved materials, spin torque equations should be independent of the extrinsic curvature.  
Similarly, the proposed geometric control of spin transport \cite{Ying2016a}, the effect of SOC in curved superconductors \cite{RevModPhys.96.021003}, and how curvature modulates the local superconducting order parameter \cite{Ying2016b} should be evaluated in light of these new insights.

On the other hand, apart from the geometrical potential $\mathcal{V}_g$, we find a novel scalar geometrical potential depending on the Rashba SOC strength, $\bar{\mathcal{W}}^B_{3}\bar{\mathcal{W}}^{B3}$, which has been unnoticed in previous studies. This SOC-induced geometrical potential, representing the remaining contribution from the normal component of the non-Abelian gauge field, does not distinguish between spins. Interestingly, this scalar potential is also present in one-dimensional systems, as can be found in Appendix \ref{AppendDecou}. 
\section{Controllability of the SDC} 
%%%%%%%%%%%%%%%%%%%%%%%%%%%%%%%%
We now turn the attention to the possibility of a induced SDC based on the analogy with an electromagnetic field in a deformed surface. For a spinless particle confined to a (flat) quasi-two-dimensional in the presence of a tilted magnetic field, one would expect that a tilted magnetic field affects the in-plane motion only via its perpendicular component ($B_{\perp}$) and no for the in-plane component ($B_{||}$), as can be verified experimentally. This must be so while the magnetic length $ \ell_{B_{||}}=\sqrt{\hbar/(e B_{||})}$ be much greater than the confinement width $\ell_c=\epsilon\sqrt{2 \hbar/(m \omega)}$.
Nevertheless, if $ \ell_{B_{||}} \sim \ell_c $ then $B_{||}$ can modified the surface energy spectrum because the spatial confinement and the magnetic field compete. Indeed, the in-plane magnetic field can lead to fictitious effects if one does not deal with terms analogous to $\mathcal{H}^{(1)}_{e}$ properly (see \cite{Brandt2018a} for more details).

On the other hand, in a deformed surface, only when there is a strong surface confinement do \textit{i}) the normal and surface states are decoupled and \textit{ii}) the SOC is unable to promote transitions between normal states. The premise of strong surface confinement, typically assumed in investigations of the SDC \cite{Chang2013a,Wang2017a,Taichi2011,Siu2018,Zhang2007a,Liu2011a,Ortix2015ab,Gentile2015a,Ying2016a,Pandey2018a,Salamone2022a}, together with the proper description of the gauge invariance leads to the vanishing of the extrinsic SDC. However, the absence of terms proportional to $\xi\text{H}\tilde{\sigma}^i$ does not imply the absence of the Rashba effect on a curved 2D surface. The Rashba SOC is intrinsically included in the covariant derivative $\mathcal{D}_{\mu}$. Therefore, this offers a potential route to use the intrinsic SDC. Additionally, our description suggests that an extrinsic induced SDC could be obtained in materials with a large enough deformation or weak confinement, where the thin-layer approximation must be refined. However, in such a scenario, transitions between normal states become allowed, and consequently, the Hilbert space of the 2D states is mixed with normal dynamics. To illustrate these results in the context of a specific Hamiltonian, we examine below the example of Gaussian bump deformations in 2D materials. Additionally, to establish a framework for realizing the SDC with a controllable deformation (e.g., the Gaussian bumps), we first design material candidates that can potentially exhibit a large SDC.  

The potential of 2D materials for hosting curvature geometry offers opportunities for advanced applications in electronics and optoelectronics~\cite{curvature_graphene}. To name a few examples, graphene, with its intrinsic elastic ripples and ability to form fold structures, demonstrates enhanced carrier mobility and modified electronic properties under curvature~\cite{curvature_graphene}, silicene~\cite{curvature_silicene} and stanene~\cite{curvature_stanene}, with buckled structures due to longer bond lengths, exhibit the quantum spin Hall effect facilitated by their geometric configuration, making them promising candidates for curvature applications. Transition metal dichalcogenides (TMDs) like MoS$_2$ and WSe$_2$ exhibit significant changes in their electronic, optical, and mechanical properties when subjected to curvature. Folded MoS$_2$, for instance, displays increased carrier mobility and bandgap modulation~\cite{curvature_tmds}, while boron nitride (BN) heterostructures, designed with atomically constructed curved structures, influence their optoelectronic properties~\cite{curvature_bn}. Black phosphorus shows changes in photoluminescence spectra under different strains induced by curvature~\cite{curvature_bp}. These materials exemplify the diverse potential for utilizing curvature geometry to enhance and control their properties in cutting-edge technological applications~\cite{curvature_graphene}.

These synthesized materials hosting curvatures and deformation are candidates to observe the consequences of the SDC mediated by the spin-orbit coupling in potential spintronics applications. Naturally, materials containing atoms with a considerable high atomic number, such as TMDs and stanene are among the most promising 2D materials for the potential spin control based on extrinsic deformations.

Another approach to identifying potential candidates for SDC realization involves a bottom-up materials design strategy. Specifically, in a previous study using data from the C2DB~\cite{c2db_repository}, we filtered a list of 358 2D materials exhibiting spin splitting near the band edges, including approximately 100 polar compounds with Rashba spin splitting. From this list, we selected materials with an energy above the convex hull smaller than 25 meV (i.e., room temperature), indicating a tendency for thermodynamic stability. 
To provide a context for the reader, the \textit{energy above the convex hull} is defined as the difference in energy between the 2D compound in question and the most stable linear combination of compounds (the convex hull) that have the same overall composition. 
A positive value of $\Delta E_{hull}$ indicates that the compound is metastable with respect to decomposition into the set of reference compounds, whereas a value of zero indicates that the compound is on the convex hull and is therefore thermodynamically stable. 
In table 1, we include the list of 17 2D materials having a significant large Rashba spin spliting (e.g., in a range of 76 meV to 373 meV) at the valence band, which can be a convenient platform for strong SDC. Indeed, besides the composition in the first column, space group symmetry, and bandgap, we also provide information for the spin splitting, including values for the SDC scalar geometrical potential depending on the Rashba SOC strength, $\bar{\mathcal{W}}_{3}\bar{\mathcal{W}}^{3}$
. These values fall within an energy range of 50 meV to 500 meV, indicating that the SDC is significantly more relevant than other perturbations like the SOC and the magnetic spin splitting which typically occur in the range of a few meV.

\begin{table}[h!]
    \centering
    \caption{List of SDC and Rashba spin splitting at the valence bands for 2D materials with polar structure and energy above the convex hull smaller than 25 meV from the C2DB Database \cite{c2db_repository}. \textit{PGS} represents the point group symmetry of the material's structure. $\Delta E_{hull}$ (meV) is the energy above convex hull reported by the C2DB database. \textit{$E_{g}$}, \textit{k-path}, $\alpha_R$, \textit{SS} and SDC stand for the energy band gap (eV), k-path between high-symmetry k-points where the spin splitting is identified, Rashba coefficient [eV/\AA$^{-1}$], spin splitting magnitude $\Delta_{ss}$ (meV) and spin-deformation coupling energy contribution from the scalar potential $\bar{\mathcal{W}}_{3}\bar{\mathcal{W}}^{3}$ (meV), respectively. }
    \begin{tabular}{l l l l l l l l}
        \hline
        \textbf{Material} & \textbf{PGS} & \textbf{$E_{\text{hull}}$} & \textbf{$E_{g}$} & \textbf{k-path} & \textbf{$\alpha_R$} & $\Delta_{ss}$  & \textbf{$\mathcal{\phi}$} \\
        \hline
        BrSbTe  & P3m1 & 0 & 1.09 &   $\Gamma$$\rightarrow$M & 2.63 & 90 & 453 \\
        & & & & $\Gamma$$\rightarrow$K & 2.45 & 108 & 394 \\
        \hline
        ISbTe  & P3m1 & 0 & 0.89 &   $\Gamma$$\rightarrow$M & 3.66 & 76 & 876 \\
        & & & & $\Gamma$$\rightarrow$K & 3.53 & 81 & 814 \\
        \hline
        ClSbSe & P3m1 & 14 & 1.18 &   $\Gamma$$\rightarrow$M & 2.68 & 93 & 471 \\
        & & & & $\Gamma$$\rightarrow$K & 2.45 & 94 & 393 \\
        \hline
        ClSbTe  & P3m1 & 8 & 1.29 &   M$\rightarrow$$\Gamma$ & 2.89 & 127 & 547 \\
        & & & & M$\rightarrow$K & 2.35 & 149 & 362 \\
        \hline
        Mo$_2$W$_2$Se$_8$  & P1 & 0 & 1.29 &   X$\rightarrow$$\Gamma$ & 1.78 & 313 & 208 \\
        \hline
        WCr$_3$S$_8$ & Pmm2 & 9 & 0.89 &   X$\rightarrow$$\Gamma$ & 1.32 & 119 & 114 \\
        \hline
        WMo$_3$Te$_8$ & P1 & 5 & 0.92 &   X$\rightarrow$$\Gamma$ & 1.50 & 234 & 148 \\
        & & & & $\Gamma$$\rightarrow$S & 1.29 & 210 & 109 \\
        \hline
        WCr$_3$Se$_8$ & P1 & 9 & 0.70 &   X$\rightarrow$$\Gamma$ & 1.16 & 140 & 088 \\
        \hline
        MoW$_3$Se$_8$  & Pm & 0 & 1.28 &   X$\rightarrow$$\Gamma$ & 2.10 & 373 & 288 \\
        & & & & $\Gamma$$\rightarrow$S & 1.24 & 236 & 101 \\
        \hline
        Cr$_2$Mo$_2$S$_8$  & Pma2 & 17 & 1.04 &   X$\rightarrow$$\Gamma$ & 1.34 & 95 & 118 \\
        & & & & Y$\rightarrow$S & 0.84 & 80 & 46 \\
        \hline
        Cr$_2$W$_2$S$_8$  & Pma2 & 14 & 0.97 &   X$\rightarrow$$\Gamma$ & 1.59 & 179 & 166 \\
        & & & & Y$\rightarrow$S & 1.10 & 88 & 80 \\
        \hline
        WCr$_3$Te$_8$  & P1 & 1 & 0.86 &   X$\rightarrow$$\Gamma$ & 1.39 & 84 & 127 \\
        \hline
        WMo$_3$Se$_8$  & P1 & 0 & 1.59 &   X$\rightarrow$$\Gamma$ & 1.82 & 214 & 216 \\
        \hline
        Cr$_2$W$_2$Se$_8$  & P1 & 15 & 0.78 &   X$\rightarrow$$\Gamma$ & 1.39 & 204 & 127 \\
        & & & & Y$\rightarrow$S & 1.32 & 128 & 114 \\
        \hline
        WMo$_3$S$_8$  & P1 & 0 & 1.58 &   X$\rightarrow$$\Gamma$ & 1.82 & 214 & 216 \\
        \hline
        MoCr$_3$Se$_8$ & Pm & 1 & 0.73 &   X$\rightarrow$$\Gamma$ & 0.98 & 103 & 63 \\
        \hline
        AsClSe & P3m1 & 13 & 1.36 &   $\Gamma$$\rightarrow$M & 3.72 & 97 & 905 \\
        & & & & $\Gamma$$\rightarrow$K & 3.48 & 101 & 794 \\
        \hline
    \end{tabular}
    \label{tab:materials_propertiesII}
\end{table}

In the band structure of a given two-dimensional material, the Rashba parameter at a specific band edge varies along different high-symmetry $k$-paths in the Brillouin zone. Specifically, for the same compound, we may observe multiple entries in Table I. Similar to the $g$-factor in the Zeeman effect, the SDC can differ for each band and $k$-symmetry path, highlighting its dependence on both real and momentum space. It is important to note that, for illustrative purposes, the Rashba parameter is assumed to be independent of spatial coordinates in order to provide approximate values for the scalar potential generated by the SDC. A detailed study and simulations of 2D materials~\cite{Yadav2023-kk} and the spacial dependence of the Rashba parameter is required to give a complete description. However, we can anticipate that the SDC scalar potential could induce energy shifts as large as the bandgap of the material, which can potentially invert the band order for compounds with SDC at both band edges. In Appendix, a list of compounds potentially having SDC simultaneously at the valence and conduction band is presented.

In 2D materials, the SDC results from own geometric deformations. In this sense, SDC can be induced when a material is in contact with a curved surface or substrate that causes local curvature changes \cite{uti2019}. However, this coupling is limited to the intrinsic properties of the material and is not transferable between different materials. This is distinct from phenomena like proximity-induced SOC, where SOC properties can be inherited from neighboring materials. In the case of SDC, the effect remains localized to the material's own curvature and is not transferred across an interface. For a more detailed discussion on curvature-induced effects, see Ref. \cite{nature_2016_curved_graphene}. 
On the other hand, it is well established that interfaces can induce the Rashba effect by breaking inversion symmetry, thereby giving rise to spin splitting even in materials that originally lack this property. This mechanism opens the possibility for inducing SDC in non-polar materials. By introducing an interface with strong SOC properties, the breaking of inversion symmetry at the boundary can trigger spin-textured states, potentially promoting SDC. Several recent studies highlight how Rashba SOC can emerge at interfaces and how non-polar materials can inherit spin-dependent properties from their proximate neighbors. For instance, Ref. \cite{prl_2021_rashba_interface} demonstrates how the Rashba effect is enhanced at metal-semiconductor interfaces, potentially enabling novel spintronic applications.

%%%%%%%%%%%%%%%%%%%%%%%%%%%%%%%%%%%%%%%
\section{The Gaussian bumps deformation} 
%%%%%%%%%%%%%%%%%%%%%%%%%%%%%%%%%%%%%%%
\begin{figure}[h]
\centering
\includegraphics[width=8.7cm]{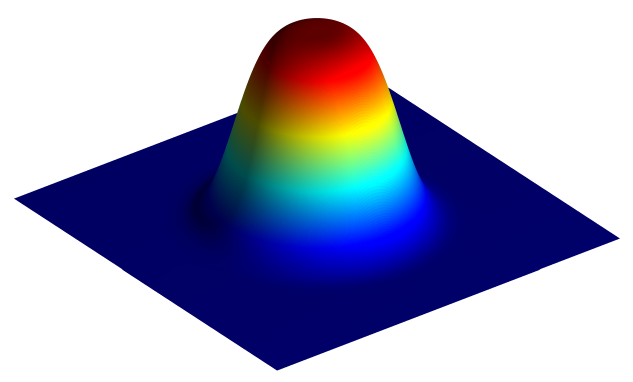}
\label{bumd}
\caption{An illustrative example showcasing a specific geometry: the Gaussian bump.}
\end{figure}

One of the most common deformation in two-dimensional materials is the Gaussian bumps located in the center of the surface \cite{PhysRevLett.87.178305,Cortijo2007a,PhysRevB.102.075425}.
Following the approach described earlier, we now focus on the study of the SDC for the curved Gaussian surfaces in typical nanostructures. A point in the surface can be represented by 
\begin{equation}
\boldsymbol{r}(r,\varphi)=r \cos \varphi \boldsymbol{\hat{i}} + r \sin \varphi \boldsymbol{\hat{j}} + f(r)\boldsymbol{\hat{k}}, 
\end{equation}
where \( r \) is the radial distance from the origin, \( \varphi \) is the polar angle measured from the positive \( x \)-axis, and \( f(r) \) represents the height above the \( xy \)-plane. The surface is curved if and only if $\dot{f}=df/dr \neq 0$. For this geometry, the metric tensor is given by $
g_{\mu \nu}=\text{diag}\left(1+\dot{f}^{2},r^{2}\right)$.
For detailed calculations of this geometry see Appendix \ref{AppenGaussianbump}. 
%If one assumes that the 3D Rashba Hamiltonian given by $H_{\mathrm{R}}= \frac{\alpha}{\hbar}\left(\sigma^x p^y-\sigma^y p^x\right)$.
Assuming the electric field $\mathbf{E} = -\nabla V$ has only a $z$ component, the non-zero components of the non-Abelian field are 
\begin{equation}
\mathcal{W}_x=-\frac{m\alpha}{\hbar} \sigma^y, \ \ \ \  \mathcal{W}_y=\frac{m\alpha}{\hbar} \sigma^x.    
\end{equation}
Using Eqs. (\ref{Fieldwr}), (\ref{Fieldwphi}), and (\ref{Fieldwq3}), the effective Hamiltonian can be obtained
\begin{equation}
\mathcal{H}_{e} =-\frac{\hbar^2}{2m}\left[\Omega_{0} +
\bar{\mathcal{W}}_{r}\Omega_{r} + \bar{\mathcal{W}}_{\varphi}\Omega_{\varphi}  \right] + \mathcal{V}_g-\mathcal{\phi}.
\end{equation}
Where we write the kinetic energy as
\begin{equation}
\Omega_{0}=g_{m}^{-1}\partial_r \left(g_{d}^{-1}\partial_r\right)
+g_{m}^{-1}\partial_{\varphi} \left(g_{d}\partial_{\varphi} \right).    
\end{equation}
Here, the notation $g_{1}=1+\dot{f}^{2}$, $g_{d}=\sqrt{g_{1}}/r$, and $g_{m}=\sqrt{g_{1}}r$ is introduced. Additionally, considering the non-Abelian gauge field as  
\begin{eqnarray}
\bar{\mathcal{W}}_{r}&=&-\frac{m\alpha}{\hbar}\cos (\varphi )\sigma^y+\frac{m\alpha}{\hbar}\sin (\varphi )\sigma^x,\\ \bar{\mathcal{W}}_{\varphi}&=&\frac{m\alpha}{\hbar}r\sin (\varphi )\sigma^y+\frac{m\alpha}{\hbar}r\cos (\varphi )\sigma^x,    
\end{eqnarray}
with $\Omega_{\varphi}=-\frac{2i}{\hbar r^2}\partial_{\varphi}$ and 
\begin{equation}
\Omega_{r}=-\frac{i}{\hbar rg_{1}}( 2r\partial_r  -g_{1} +1-r \dot{f} \ddot{f}/g_{1}).    
\end{equation}
Observe that the dependence of \(\Omega_r\) on \(f\), which characterizes the curvature, results in spatially varying spin-orbit coupling on curved surfaces. Finally, the term $\mathcal{V}_g-\mathcal{\phi}$ is given by
\begin{equation}
\mathcal{V}_g-\mathcal{\phi}=-\frac{\hbar^2 \left(-r \ddot{f}+\dot{f}^3+\dot{f}\right)^2}{8 m r^{2} g_{1}^{3}}-\frac{m\alpha^2}{2\hbar^2}\frac{\dot{f}^2}{g_{1}},
\end{equation}
The first term on the right side corresponds to the geometric potential, which depends on both the intrinsic and extrinsic geometry, whereas the second term corresponds to a (curvature-induced) scalar potential. 

\begin{figure}[h]
\centering
\includegraphics[width=8.0cm]{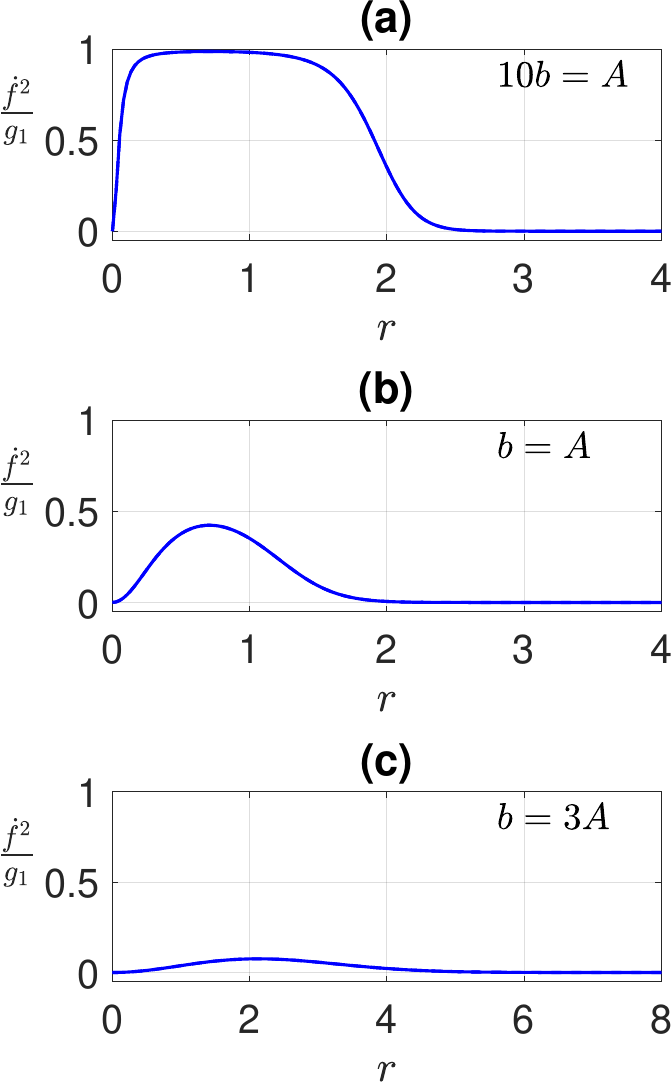}
\caption{We illustrate the behavior of function $\dot{f}^2/g_{1}$ as a function of $r$ for three parameter configurations: (a) $10b=A$, (b) $b=A$, and (c) $b=3A$. The plots show that as the ratio of parameter $A$ to parameter $b$ increases, the potential well, given by $-\frac{m\alpha^2}{2\hbar^2}\frac{\dot{f}^2}{g_{1}}$ becomes deeper and flatter.}
\label{MongeRadial22}
\end{figure}

We now examine the final term. It is consistently negative and independent of the extrinsic curvature. In the specific case of a Gaussian bump described by \( f(r) = A \exp(-r^2/b^2) \), the term reaches its minimum at \( r = b/\sqrt{2} \), where \( \frac{\dot{f}^2}{g_{1}} \) reaches its maximum value. This maximum value is given by \(\left(1 + \frac{e b^2}{2 A^2} \right)^{-1} \) which is always less than one. Fig. \ref{MongeRadial22} show the function \( \frac{\dot{f}^2}{g_{1}} \) for three parameter sets.

This example illustrates how the curvature can influence the spin dynamics of particles through derivatives of the metric tensor, indicating the presence of intrinsic spin-dependent coupling.

\section{Conclusion} 
%In devices applications, one of the biggest challenges is always finding a mechanism to control a property. Although these mechanisms have been extensively studied from an electronic point of view, when we refer to the degree of freedom of the spin, it is an area that is still developing. 
%Different approaches and results explore the possibility of controlling the spin splitting in the electronic band structure of 2D material  or thin-layers through deformations as a potential route for the realization of devices, adding to the family of SOC and spin-valley among others, the SDC. 
Based on the unitary transformation usually used in many-body physics for the separation of variables and asymptotic behavior, we derive a general methodology for the separation of normal and surface degrees of freedom. 
%Although there are no geometrical reasons to \textit{a priori} neglect the effects of extrinsic curvature in the effective SOC Hamiltonian, 
We demonstrate that, due to gauge invariance, the SDC  depends only on the intrinsic geometry of the surface. 
Therefore, a surface with zero Gaussian curvature does not modify the spin dynamics.
We have also find previously unnoticed scalar geometrical potential proportional to the Rashba SOC strength. 
In the proposed methodology, it is intrinsically assumed that the non-Abelian gauge field describing the SOC acts as an external field, implying that the components of $\mathcal{W}_i$ ($i=1,2,3$) remain unaltered when the system suffers deformations.
This assumption deserves future considerations. 
In Appendix \ref{AppendDecou}, we provide a detailed general deduction applicable to both quantum surfaces and wires. %Consequently, the aforementioned conclusions are also transferable to case of quantum wires.  
Additionally, our work suggests that a route to induce SDC is to consider distortions in the crystalographic point group of three-dimensional materials, as usually demonstrated by density functional calculations. The use of the proposed formalism for the study of SDC in the context of other possible deformations can open a path for future work in this field.

\appendix

\section{Perturbative expansion in a general $m$-dimensional submanifold}\label{AppenAmanifold}

%%%%%%%%%%%%%%%%%%%%%%%%%%%%%%
%The $SU(2)$ non-Abelian  gauge field  satisfied the Coulomb gauge restriction, $\nabla \cdot \boldsymbol{\mathcal{W}}$ \cite{Shikakhwa_2012}. Employing $\boldsymbol{\mathcal{W}}$, the Hamiltonian can be written into the form
%\begin{equation}
%H_{e}(\boldsymbol{\mathcal{W}})=\frac{(\boldsymbol{p}-e\boldsymbol{\mathcal{A}}-\boldsymbol{\mathcal{W}})^2}{2 m}-\frac{1}{2 m} \boldsymbol{\mathcal{W}} \cdot \boldsymbol{\mathcal{W}}+\mathcal{W}_0+e\mathcal{A}_0.
%\end{equation}
%Usually, the quadratic term in the gauge field, which is spin-independent and breaks the $SU(2)$ gauge symmetry, is absorbed into a scalar potential. However, in our work, we prefer to leave this term explicit. Analogously to what was done in Refs. \cite{Ferrari2008,Brandt_2017,Brandt2015a,monroy2017teoria,Liang2020a}, we employ a gauge covariant derivative $\mathcal{D}_{i}=\partial_{i}-\frac{ie}{\hbar}\mathcal{A}_{i}-\frac{i}{\hbar}\mathcal{W}_{i}$. 

In this appendix, we address the most general scenario wherein the particle is confined to an $m$-dimensional submanifold –a low-dimensional space ($\mathcal{M}$)–within an $n$-dimensional manifold ($\mathcal{N}$). The particle interacts with an external (Abelian) electromagnetic field, which propagates into an $n$-dimensional Euclidean space, along with a non-Abelian gauge field (Rashba SOC). The procedure extends the results presented in \cite{Brandt2015a,Brandt_2017,Liang2020a}. 
%%%%%%%%%%%%%%%%%%
\subsection*{Geometrical preliminaries}\label{AppenGeome}
%%%%%%%%%%%%%%%%%%
We assume that $\mathcal{M}$ has a tubular neighborhood, such that \cite{monroy2017teoria}
\be
\boldsymbol{R}(x^{\mu},y^i)=\boldsymbol{r}(q^{\mu})+q^j\hat{\boldsymbol{n}}_j(q^{\mu}),
\ee
defines an arbitrary point in the vicinity of $\mathcal{M}$, where $\hat{\boldsymbol{n}}_j$ are orthonormal vectors to $\mathcal{M}$. If $\boldsymbol{t}_{\mu}=\frac{\partial \boldsymbol{r}}{\partial q^{\mu}}$ are $m$ tangent vectors on $\mathcal{M}$, the first fundamental form (metric), the second fundamental form (extrinsic curvature) and the normal fundamental form (geometric or extrinsic torsion) are respectively,
\be\label{metrictv}
&g_{\mu\nu}\equiv {\boldsymbol{t}}_\mu\cdot{\boldsymbol{t}}_\nu, &\\ \label{secondfundamentalform}
&\alpha_{i\mu\nu} \equiv -{\boldsymbol{ t}}_\mu\cdot\partial_\nu\hat{\boldsymbol{n}}_i, &\\
&\beta_{\mu i}^{\ \ k} \eta_{kj} \equiv\hat{\boldsymbol{n}}_i\cdot\partial_\mu \hat{\boldsymbol{n}}_j&.
\ee
Note that extrinsic torsion arises only when the codimension is greater than one, $n-m>1$. Previously, it was shown that any torsion contributes solely to coupling with angular momentum operators in normal space  \cite{Jaffe2003,monroy2017teoria}. For confining potentials exhibiting $SO(p)$ symmetry, the eigenstates of the effective Hamiltonian $\mathcal{H}^{(0)}_{e}$ can be decomposed into irreducible $SO(p)$ multiplets: Consequently, if the system is initially prepared in the ground state of $\mathcal{H}^{(0)}_{e}$, its angular momentum in the normal coordinates vanishes, and considering torsion will add nothing. Thus, for the sake of simplicity, in the subsequent discussion, we assume that the manifold is devoid of extrinsic torsion. Therefore, the metric for the Euclidean space is given by:
\be
G_{AB} =\left(\begin{array}{ll}
\gamma_{\mu\nu}&
      0 \\
 0  & \delta_{ij}
\end{array}\right)
\ee
%\be
%G_{AB} =\left(\begin{array}{ll}
%\gamma_{\mu\nu}+q^k q^l \beta_{\mu k}^{\ \ h} \beta_{\nu l}^{\ \ m}\delta_{hm}&
%       -q^k \beta_{\mu k}^{\ \ h}\delta_{ih} \\
% -q^k \beta_{\nu k}^{\ \ h}\delta_{jh}  & \delta_{ij}
%\end{array}\right)
%\ee
where $\gamma_{\mu \nu}$ is given by
\be
\gamma_{\mu \nu}=g_{\mu \nu}-2q^{k}\alpha_{k \mu \nu}+q^{k}q^{l}\alpha_{k \mu\rho}g^{\rho\sigma}\alpha_{l \sigma \nu}.
\ee
Here, we employ Latin indices $A$, $B$ running from $1$ to $n$ for the coordinates in the Euclidean space, Latin indices $i$, $j, \dots$ run from $1$ to $n-m$ and the Greek indices $\mu$, $\nu, \dots$  run from $1$ to $m$. Denoting the determinants of $G_{AB}$ and $\gamma_{\mu \nu}$ as $|G|$ and $|\gamma|$, respectively, it is possible to show that $|G|=|\gamma|$. The inverse of the metric tensor $G_{AB}$ can be calculated exactly, and is given by
\be\label{metricinversenm}
G^{AB}= \left(\begin{array}{ll}
\lambda^{\mu \nu} & 0\\
0 & \delta^{ij}\end{array}\right),
\ee
%\be\label{metricinversenm}
%G^{AB}= \left(\begin{array}{ll}
%\lambda^{\mu \nu} & \lambda^{\mu\sigma}q^{k}\beta_{\sigma k}^{\ \ j}\\
%\lambda^{\nu\sigma}q^{k}\beta_{\sigma k}^{\ \ i} & \delta^{ij}+q^{k}q^{l}\beta_{\sigma k}^{\ \ i}\beta_{\rho l}^{\ \ j}\lambda^{\sigma\rho}\end{array}\right),
%\ee
where $\lambda^{\mu \nu}\equiv(\gamma^{-1})_{\mu \nu}$ is the inverse of $\gamma_{\mu \nu}$ \cite{Jaffe2003}. 
%%%%%%%%%%%%%%%%%%
\subsection*{Perturbative expansion}\label{AppenPertur}
%%%%%%%%%%%%%%%%%%
We rescale the normal coordinates by $q \rightarrow \epsilon q$. Expanding $\mathcal{H}_{e}$ in powers of $\epsilon$ yields
\be
\epsilon^2\mathcal{H}_{e}=\mathcal{H}^{(0)}_{e}+\epsilon \mathcal{H}_{e}^{(1)}+\epsilon^2 \mathcal{H}^{(2)}_{e}+\ldots .
\ee
As $\boldsymbol{\mathcal{Z}}(q^{\mu},\epsilon q^i)$ and $\boldsymbol{\mathcal{W}}(q^{\mu},\epsilon q^i)$ are now dependent on $\epsilon$ their Taylor expansions are given by Eqs. (\ref{boldsymbolz}) and (\ref{boldsymbolw}), respectively. Therefore, the contributions to the zero, first and second order terms are
\be
\mathcal{H}^{(0)}_{e}&=&-\frac{\hbar^2}{2m}\partial_i\partial^i +V_c,\\\label{FirstCoupling1}
\mathcal{H}^{(1)}_{e}&=&-\frac{i}{m}\hbar\bar{\mathcal{Z}}_{i}\partial^{i},\\\nonumber
\ee

\begin{widetext}
\be\nonumber
2m\mathcal{H}^{(2)}_{e}\chi&=&
-\frac{\hbar^2}{|g|^{1/2}}\bigg[\partial_{\mu}g^{\mu\nu}|g|^{1/2}\partial_{\nu}\bigg]\chi+2\mathcal{V}_g m\chi
-2i\hbar \bar{\mathcal{Z}}_{\mu}g^{\mu \nu} \partial_{\nu}\chi-\frac{i\hbar}{|g|^{1/2}}\partial_{\mu}\left(|g|^{1/2}g^{\mu \nu} \bar{\mathcal{Z}}_{\nu}\right)\chi
\\\nonumber\label{effectivebeforegt}
&&
-(\bar{\mathcal{Z}}_i \bar{\mathcal{Z}}^i+\bar{\mathcal{Z}}_{\mu}g^{\mu\nu}\bar{\mathcal{Z}}_{\nu})\chi
- 2i\hbar \bar{\mathcal{Z}}_i q^i H \chi
-i\hbar\frac{\partial  \bar{\mathcal{Z}}_i}{\partial q_i}\chi-2i\hbar \frac{\partial  \bar{\mathcal{Z}}_i}{\partial q^j} q^j\partial^{i}\chi
\\
\label{SecondOrder1}
&&-\left[ \bar{\mathcal{W}}_{i}\bar{\mathcal{W}}^{i}+\bar{\mathcal{W}}_{\mu}g^{\mu \nu} \bar{\mathcal{W}}_{\nu}\right]\chi
+2m\mathcal{Z}_0\chi,
\ee
\end{widetext}
where $\mathcal{V}_g$ is a \textit{geometric potential} given by
\be\label{Geometricalpotentialm}
V_G=-\frac{\hbar^2}{8m}g^{\nu\mu}g^{\rho\sigma}\left(\alpha_{k\mu\nu}\alpha^k_{\ \rho\sigma}
-2\alpha_{k\mu\sigma}\alpha^k_{\ \rho\nu}\right). 
\ee
%%%%%%%%%%%%%%%%%%%%%

\section{Decoupling in the Abelian case}\label{AppendDecou}
%%%%%%%%%%%%%%%%%%
The dynamics of the tangent and normal degrees of freedom are coupled by the first-order term (Eq. (\ref{FirstCoupling1})), whereas the curvature is coupled with the components of fields $\bar{\mathcal{W}}_i$ and $\bar{\mathcal{A}}_i$ in the sixth term of (Eq. (\ref{SecondOrder1})). In addition, the in-plane and out-of-plane modes are coupled by the seventh and eighth terms of (Eq. (\ref{SecondOrder1})).

To obtain the effective dynamics, it is necessary to ``freeze'' the normal degrees of freedom \cite{Maraner95,Jaffe2003}. However, the first-order term (Eq. (\ref{FirstCoupling1})) couples the dynamics of the tangent and normal degrees of freedom through the normal components of the $U(1)\times SU(2)$ gauge potential ($\boldsymbol{\mathcal{Z}}$) evaluated at $q^j$ equal to zero ($\bar{\mathcal{Z}}_i=\mathcal{Z}_i(q^{\mu},0)$). In the case that there was only an Abelian field, the dynamics could be decoupled by performing a gauge transformation \cite{Brandt2015a,monroy2017teoria,Brandt_2017}
\begin{equation}\label{Abeliangaugetransform}
\gamma(x^\mu,\epsilon q^i)= -\epsilon \bar{\mathcal{A}}_i q^i-\frac{\epsilon^2}{2}\frac{\partial  \bar{\mathcal{A}}_i}{\partial q^j} q^i q^j+\mathcal{O}((\epsilon q)^3),
\end{equation}
%\begin{equation}
%\gamma(x^\mu,\epsilon q^i)=-\int_0^{q^j}\epsilon %\mathcal{A}_j(x^\mu,\epsilon q^i)dq^j,
%\end{equation}
for which it is satisfied that
\begin{eqnarray}
\bar{\mathcal{A}}'_i=\bar{\mathcal{A}}_{i}+\frac{1}{\epsilon}\partial_i \gamma'=-\epsilon \frac{q^j}{2}\bar{\mathcal{F}}_{ij}+\mathcal{O}((\epsilon q)^2),
\end{eqnarray}
where $\bar{\mathcal{F}}_{ij}=\left.\mathcal{F}_{ij}\right|_{q=0}$
is the (Abelian) electromagnetic field tensor. 

For the non-Abelian case, we proceed similarly, using the gauge transformation (\ref{nonAbeliangaugetransform}) to decouple the dynamics. After performing both Abelian and non-Abelian gauge transformations, the zero, first, and second order terms are as follows:
\begin{equation}
\mathcal{H}^{(0)}_{e}=-\frac{\hbar^2}{2m}\partial_3\partial^3 +\mathcal{V}_c,
\end{equation}
\begin{equation}
\mathcal{H}^{(1)}_{e}=0,
\end{equation}
and
\begin{widetext}
\begin{eqnarray}
2m\mathcal{H}^{(2)}_{e}\chi&=&
-\frac{\hbar^2}{|g|^{1/2}}\bigg[\partial_{\mu}g^{\mu\nu}|g|^{1/2}\partial_{\nu}\bigg]
\chi+2\mathcal{V}_g m\chi
-2i\hbar \bar{\mathcal{Z}}_{\mu}g^{\mu \nu} \partial_{\nu}\chi-\frac{i\hbar}{|g|^{1/2}}\partial_{\mu}\left(|g|^{1/2}g^{\mu \nu} \bar{\mathcal{Z}}_{\nu}\right)\chi
\\
&&
-\bar{\mathcal{Z}}_{\mu}g^{\mu\nu}\bar{\mathcal{Z}}_{\nu}\chi
-\left[ \bar{\mathcal{W}}^B_{i}\bar{\mathcal{W}}^{Bi}+\bar{\mathcal{W}}_{\mu}g^{\mu \nu} \bar{\mathcal{W}}_{\nu}\right]\chi
+2m\mathcal{Z}_0\chi +e\bar{\mathcal{F}}^{ij}L_{ij} \chi+\bar{\mathcal{G}}^{ij}L_{ij}\chi.
\end{eqnarray}
\end{widetext}
Here, $L_{ij}=i\hbar(y_{j}\partial_{i}-y_{i}\partial_{j})$ are the angular momentum operators in the space normal to $\mathcal{M}$. As before, if the system is initially prepared in the ground state of $\mathcal{H}^{(0)}_{e}$, its angular momentum in the normal coordinates vanishes, and the last two terms do not contribute. Considering the above and in the case of a surface, this directly leads to Eq. (\ref{FinalEffectiveRashba1}).
%Finally, note that under the Abelian gauge transformation, the wave function $\Psi$ is transformed as $\Psi' = e^{\frac{ie\gamma}{\hbar}}\Psi$.
%%%%%%%%%%%%%%%%%%%%%%%%%

\section{List of SDC and Rashba spin splitting at the conduction bands for 2D materials}\label{ListOfMaterials}

\begin{table}[h!]
    \centering
    \caption{List of SDC and Rashba spin splitting at the conduction bands for 2D materials with polar structure and energy above the convex hull smaller than 25 meV from the C2DB Database \cite{c2db_repository}. }
    \begin{tabular}{l l l l l l l l}
        \hline
       \textbf{Material} & \textbf{PGS} & \textbf{$E_{\text{hull}}$} & \textbf{$E_{g}$} & \textbf{k-path} & \textbf{$\alpha_R$} & $\Delta_{ss}$  & \textbf{$\mathcal{\phi}$} \\
        \hline
        SSeW  & P3m1 & 0.01 & 1.417 & $\Gamma$$\rightarrow$M & 3.284 & 177 & 0.706 \\
        & & & & $\Gamma$$\rightarrow$K & 3.288 & 214 & 0.708 \\
        \hline
        BiBrS & P3m1 & 0.0 & 1.227 & $\Gamma$$\rightarrow$M & 1.271 & 164 & 0.106 \\
        & & & & $\Gamma$$\rightarrow$K & 1.293 & 162 & 0.11 \\
        \hline
        BiClS & P3m1 & 0.12 & 1.841 & M$\rightarrow$$\Gamma$ & 0.461 & 224 & 0.014 \\
        \hline
        BiITe & P3m1 & 0.0 & 0.701 & $\Gamma$$\rightarrow$M & 2.065 & 128 & 0.279 \\
        & & & & M$\rightarrow$K & 1.391 & 137 & 0.127 \\
        \hline
        & & & & $\Gamma$$\rightarrow$K & 2.086 & 127 & 0.285 \\
        \hline
        ISbSe & P3m1 & 0.13 & 1.078 & $\Gamma$$\rightarrow$M & 1.17 & 101 & 0.09 \\
        & & & & $\Gamma$$\rightarrow$K & 0.954 & 122 & 0.06 \\
        \hline
        Bi2P2Te6 & P1 & 0.14 & 0.507 & $\Gamma$$\rightarrow$Y & 1.814 & 139 & 0.216 \\
        & & & & Y$\rightarrow$$\Gamma$ & 1.725 & 91 & 0.195 \\
        \hline
        & & & & X$\rightarrow$$\Gamma$ & 1.692 & 91 & 0.188 \\
        \hline
        & & & & $\Gamma$$\rightarrow$X & 1.809 & 133 & 0.214 \\
        \hline
        BiIS & P3m1 & 0.14 & 0.848 & $\Gamma$$\rightarrow$M & 1.224 & 144 & 0.098 \\
        & & & & $\Gamma$$\rightarrow$K & 1.254 & 141 & 0.103 \\
        \hline
        BiITe & P3m1 & 0.11 & 0.691 & M$\rightarrow$$\Gamma$ & 0.467 & 132 & 0.014 \\
        \hline
        BiClS & P3m1 & 0.0 & 1.334 & $\Gamma$$\rightarrow$M & 1.146 & 115 & 0.086 \\
        & & & & $\Gamma$$\rightarrow$K & 1.181 & 112 & 0.091 \\
        \hline
    \end{tabular}
    \label{tab:materials_properties}
\end{table}

\begin{table}[h!]
\centering
   \caption{List of SDC and Rashba spin splitting at both conduction (C) and valence (V) bands for 2D materials with polar structure and energy above the convex hull smaller than 25 meV from the C2DB Database \cite{c2db_repository}.  }
\begin{tabular}{l l l l l l l l l}
\hline\hline
        \textbf{Material} & \textbf{PGS} & \textbf{$E_{\text{hull}}$} & \textbf{$E_{g}$} & Band & \textbf{k-path} & \textbf{$\alpha_R$} & $\Delta_{ss}$  & \textbf{$\mathcal{\phi}$} \\
\hline
ISbSe & P3m1 & 0 & 1.061 & V & M$\rightarrow$K & 2.431 & 146 & 387 \\
 &  &  &  & C & $\Gamma$$\rightarrow$M & 1.589 & 156 & 165 \\
 &  &  &  & C & $\Gamma$$\rightarrow$K & 1.616 & 154 & 171 \\
\hline
BiBrSe & P3m1 & 0 & 1.03 & V & $\Gamma$$\rightarrow$M & 2.784 & 80 & 508 \\
 &  &  &  & V & $\Gamma$$\rightarrow$K & 2.634 & 82 & 454 \\
 &  &  &  & C & $\Gamma$$\rightarrow$M & 1.366 & 140 & 122 \\
 &  &  &  & C & $\Gamma$$\rightarrow$K & 1.417 & 137 & 132 \\
\hline
BiClTe & P3m1 & 0 & 0.938 & V & M$\rightarrow$K & 1.175 & 208 & 9 \\
 &  &  &  & C & M$\rightarrow$K & 0.992 & 113 & 64 \\
\hline
AsISe  & P3m1 & 0 & 1.164 & V & M$\rightarrow$K & 1.626 & 153 & 173 \\
 &  &  &  & C & $\Gamma$$\rightarrow$M & 1.583 & 116 & 164 \\
 &  &  &  & C & $\Gamma$$\rightarrow$K & 1.615 & 113 & 171 \\
\hline
BiBrTe & P3m1 & 0 & 0.878 & V & M$\rightarrow$K & 1.13 & 112 & 84 \\
 &  &  & & C & M$\rightarrow$K & 1.08 & 117 & 76 \\
\hline
MoW3S8 & P1 & 0 & 1.552 & V & X$\rightarrow$$\Gamma$ & 2.241 & 354 & 329 \\
 &  &  &  & C & S$\rightarrow$Y & 1.202 & 83 & 95 \\
\hline
BiIS & P3m1 & 14 & 1.139 & V & M$\rightarrow$K & 1.707 & 166 & 191 \\
 &  &  &  & C & $\Gamma$$\rightarrow$M & 1.858 & 266 & 226 \\
 &  &  &  & C & M$\rightarrow$K & 1.074 & 111 & 76 \\
 &  &  &  & C & $\Gamma$$\rightarrow$K & 1.645 & 297 & 177 \\
\hline
Mo2W2Te8 & Pm & 11 & 0.879 & V & X$\rightarrow$$\Gamma$ & 1.645 & 259 & 177 \\
 &  &  &  & V & $\Gamma$$\rightarrow$S & 1.315 & 217 & 113 \\
 &  &  &  & C & $\Gamma$$\rightarrow$S & 1.35 & 86 & 119 \\
\hline
BiISe & P3m1 & 0 & 0.929 & V & M$\rightarrow$K & 1.267 & 87 & 105 \\
 &  &  &  & C & $\Gamma$$\rightarrow$M & 2.045 & 232 & 274 \\
 &  &  &  & C & M$\rightarrow$K & 1.142 & 115 & 85 \\
 &  &  &  & C & $\Gamma$$\rightarrow$K & 2.089 & 228 & 286 \\
\hline
Mo2W2S8 & Pc & 0 & 1.553 & V & X$\rightarrow$$\Gamma$ & 1.954 & 278 & 250 \\
 &  &  &  & C & S$\rightarrow$Y & 1.167 & 86 & 89 \\
\hline
BiClSe & P3m1 & 0 & 1.139 & V & $\Gamma$$\rightarrow$M & 2.393 & 118 & 375 \\
 &  &  &  & V & M$\rightarrow$K & 1.782 & 90 & 208 \\
 &  &  &  & V & $\Gamma$$\rightarrow$K & 2.216 & 124 & 322 \\
 &  &  &  & C & M$\rightarrow$$\Gamma$ & 3.032 & 92 & 602 \\
\hline\hline
\end{tabular}
\label{tab:materials}
\end{table}

%%%%%%%%%%%%%%%%%%%%%%%%%
\section{Details of the Gaussian bump}\label{AppenGaussianbump}
%%%%%%%%%%%%%%%%%%%%%%%%%
To accurately represent a surface in a three-dimensional space, we employ Monge parametrization \cite{kuehnel2015}, where the surface is described by a single height function relative to a reference plane. In our case, the two coordinates used to describe the surface are the $r$ and $\varphi$ coordinates of the cylindrical coordinate system, so $\boldsymbol{r}_{||}=\boldsymbol{r}(r,\varphi)$. Any point on the surface is given by $\boldsymbol{r}(r,\varphi)=r \cos \varphi \boldsymbol{\hat{i}} + r \sin \varphi \boldsymbol{\hat{j}} + f(r)\boldsymbol{\hat{k}}$, where $r$ is the radius, $\varphi$ is the angle, and $z$ is the height. The tangent vectors are given by ${t}_{\mu}=d\boldsymbol{r}_{||}/dq^{\mu}$, so
$$
\boldsymbol{t}_{\varphi}=\frac{\partial \vec{r}}{\partial \varphi}=\left(\begin{array}{c}
-r \sin \varphi \\
r \cos \varphi \\
0
\end{array}\right) \quad \text { and } \quad \boldsymbol{t}_r=\frac{\partial \vec{r}}{\partial r}=\left(\begin{array}{c}
\cos \varphi \\
\sin \varphi \\
\dot{f}
\end{array}\right),
$$
with \(\dot{f} = \frac{df}{dr}\), the metric, or first fundamental form, on the surface \(S\) is defined by Eq. (\ref{metrictv}), in this case is given by $
g_{\mu \nu}=\text{diag}\left(1+\dot{f}^{2},r^{2}\right)$. The unit normal vector field can be determined as follows

\begin{equation}
\hat{\boldsymbol{n}} =\frac{\boldsymbol{t}_{\varphi} \times \boldsymbol{t}_{r}}{\left|\boldsymbol{t}_{\varphi} \times \boldsymbol{t}_{r} \right|}=\frac{1}{\sqrt{1+\dot{f}^2}}\left(\begin{array}{c}
\dot{f} \cos \varphi \\
\dot{f} \sin \varphi \\
-\dot{r}
\end{array}\right)
\end{equation}
We can then calculate the Weingarten curvature matrix $\alpha_{\mu \nu} \equiv -\boldsymbol {t}_{\mu}\cdot\partial_{\nu}\hat{\boldsymbol{n}}=\partial_{\nu} \boldsymbol {t}_{\mu}\cdot\hat{\boldsymbol{n}}$ is
\begin{eqnarray}
\alpha_{\mu \nu}=\frac{1}{\sqrt{1+\dot{f}^2}}\left(\begin{array}{cc}
-\ddot{f}  & 0 \\
0 & -r \dot{f} 
\end{array}\right) \text {, }    
\end{eqnarray}
so,
\begin{eqnarray}\nonumber
\alpha_{\mu}^{\nu}&=&-\alpha_{\mu \nu} g^{\mu \nu}\\\nonumber
&=&\left(\begin{array}{cc}
\frac{\ddot{f} }{\left(1+\dot{f}^2 \right)^{3 / 2}} & 0 \\
0 & \frac{\dot{f} }{r \sqrt{1+\dot{f}^2}} 
\end{array}\right)\\
&=&\left(\begin{array}{cc}
-\kappa_1 & 0 \\
0 & -\kappa_2
\end{array}\right),
\end{eqnarray}
where $\kappa_1$ and $\kappa_2$ are the principal curvatures. In general, for the Monge parameterization, we have $\boldsymbol{R}=\boldsymbol{r}_{||}+q^3\hat{\boldsymbol{n}}$ and $\hat{\boldsymbol{n}} =\frac{1}{\sqrt{1+\dot{f}^2}}(
\dot{f} \cos \varphi \boldsymbol{\hat{i}} + \dot{f} \sin \varphi \boldsymbol{\hat{j}} -\dot{r} \boldsymbol{\hat{k}})$, then  
\begin{eqnarray}\nonumber
\boldsymbol{R}(r,\varphi,q^{3})&=& \cos \varphi\left(r + q^{3}\frac{\dot{f}}{\sqrt{1+\dot{f}^2}}
\right)\boldsymbol{\hat{i}}\\\nonumber && + \sin \varphi \left(r + q^{3}\frac{\dot{f} }{\sqrt{1+\dot{f}^2}}
\right) \boldsymbol{\hat{j}}\\ &&+ \left(f(r) - q^{3}\frac{\dot{r}}{\sqrt{1+\dot{f}^2}}
\right)\boldsymbol{\hat{k}}.
\end{eqnarray}
%therefore, 
%\begin{eqnarray}
     %x&=&\cos \varphi\left(r + q^{3}\frac{\dot{f}}{\sqrt{1+\dot{f}^2}}
%\right),\\
%y&=& \sin \varphi \left(r + q^3\frac{\dot{f} }{\sqrt{1+\dot{f}^2}}
%\right),\\
%z&=&\left(f(r) - q^3\frac{\dot{r}}{\sqrt{1+\dot{f}^2}}
%\right).
%\end{eqnarray}
To obtain the $SU(2)$ non-Abelian gauge field in the adapted coordinate frame evaluated at $q^3=0$, we need to compute:
\begin{widetext}
\begin{eqnarray}
\left. \left\{\frac{\partial x}{\partial r},\frac{\partial x}{\partial \varphi},\frac{\partial x}{\partial q^3}\right\}\right|_{q^3=0}&=&\left\{\cos (\varphi ),-r\sin (\varphi ) ,\frac{\dot{f}(r) \cos (\varphi )}{\sqrt{\dot{f}(r)^2+1}}\right\},\\
\left.  \left\{\frac{\partial y}{\partial r},\frac{\partial y}{\partial \varphi},\frac{\partial y}{\partial q^3}\right\}\right|_{q^3=0}&=&\left\{\sin (\varphi ),r\cos (\varphi ) ,\frac{\dot{f}(r)\sin (\varphi )}{\sqrt{\dot{f}(r)^2+1}}\right\},\\
\left.  \left\{\frac{\partial z}{\partial r},\frac{\partial z}{\partial \varphi},\frac{\partial z}{\partial q^3}\right\}\right|_{q^3=0}&=&\left\{\dot{f}(r) \sin (\varphi ),f(r) \cos (\varphi ),-\frac{\dot{f}(r) \sin (\varphi )}{\sqrt{\dot{f}(r)^2+1}}\right\}.
\end{eqnarray}
\end{widetext}
Assuming that the electric field $\mathbf{E}=-\nabla V$ has only a $z$ component, the non-zero components of the non-Abelian field are $\mathcal{W}_x=-\frac{m\alpha}{\hbar}\sigma^y$, and $\mathcal{W}_y=\frac{m\alpha}{\hbar}\sigma^x$. Therefore, the field in the adapted coordinate frame, $q^i=(r,\varphi,q^3)$, 
can be calculated using $\mathcal{W}_{q^i}=\frac{\partial x^{\mu}}{\partial q^i}\mathcal{W}_{x^{\mu}}$, so
\begin{eqnarray}\nonumber
    \bar{\mathcal{W}}_{r}&=&\bar{\mathcal{W}}_{x}\cos (\varphi )+\bar{\mathcal{W}}_{y}\sin (\varphi )\\\label{Fieldwr}
    &=&-\frac{m\alpha}{\hbar}\cos (\varphi )\sigma^y+\frac{m\alpha}{\hbar}\sin (\varphi )\sigma^x,
\end{eqnarray}
\begin{eqnarray}\nonumber
    \bar{\mathcal{W}}_{\varphi}&=&-\bar{\mathcal{W}}_{x}r\sin (\varphi )+\bar{\mathcal{W}}_{y}r\cos (\varphi )\\\label{Fieldwphi}
    &=&\frac{m\alpha}{\hbar}r\sin (\varphi )\sigma^y+\frac{m\alpha}{\hbar}r\cos (\varphi )\sigma^x,
\end{eqnarray}
\begin{eqnarray}\nonumber
    \bar{\mathcal{W}}_{q^3}&=&\bar{\mathcal{W}}_{x}\frac{\dot{f}(r) \cos (\varphi )}{\sqrt{\dot{f}(r)^2+1}}+\bar{\mathcal{W}}_{y}\frac{\dot{f}(r) \sin (\varphi )}{\sqrt{\dot{f}(r)^2+1}}
    \\\label{Fieldwq3}
    &=&\frac{-m\alpha\dot{f}(r)}{\hbar\sqrt{\dot{f}(r)^2+1}}\left(\cos (\varphi )\sigma^y+\sin (\varphi )\sigma^x\right).
\end{eqnarray}

%%%%%%%%%%%%%%%%%%%%%%%%%
%\end{acknowledgements}
  \bibliographystyle{apsrev4-2}
%\bibliography{Rashbabiblio}%\label{refs}
\bibliography{main-Extended.bbl}%\label{refs}
%%%%%%%%%%%%%%%%%%%%%%%%%%%%%
\end{document}